\documentclass[onecolumn]{aa}
\usepackage{graphicx}
\usepackage{txfonts}
\begin{document}
   \title{On the V-type asteroids outside the Vesta family}

   \subtitle{I. Interplay of nonlinear secular resonances
and the Yarkovsky effect: the cases of 956 Elisa and 809 Lundia}

   \author{V. Carruba
          \inst{1}
	  \and
	  T. A. Michtchenko\inst{1}
	  \and
          F. Roig\inst{2}
          \and
	  S. Ferraz-Mello\inst{1}
	  \and 
	   D. Nesvorn\'{y}\inst{3} \fnmsep
          }

   \offprints{V. Carruba}

   \institute{ IAG, Universidade de S\~{a}o Paulo, S\~{a}o Paulo, 
     SP 05508-900, Brazil\\
              \email{valerio@astro.iag.usp.br}
         \and
Observat\'{o}rio Nacional, Rio de Janeiro, RJ 20921-400, Brazil
             \email{froig@on.br}
	     \and
Southwest Research Institute, Department of Space Studies,  
Boulder, Colorado 80302\\
             \email{davidn@boulder.swri.edu}
             }

   \date{Received May 2$^{nd}$ 2005; accepted June 22$^{nd}$ 2005.}

   \abstract{
Among the largest objects in the main belt, asteroid 4 Vesta is 
unique in showing a basaltic crust.  It is also the 
biggest member of the Vesta family, which is 
supposed to originate from a large cratering event about 1 Gyr 
ago (Marzari et al. 1996).  Most 
of the members of the Vesta family for which a spectral 
classification is available show a V-type spectra.  Due to their 
characteristic infrared spectrum, V-type asteroids are easily 
distinguished.  Before the discovery of 1459 Magnya (Lazzaro 
et al. 2000) and of several V-type NEA (Xu 1995), all the known 
V-type asteroids were members of the Vesta family.  
Recently two V-type asteroids, 809 Lundia and 956 Elisa, 
(Florczak et al. 2002) have been discovered well outside the limits 
of the family, near the Flora family.  We 
currently know 22 V-type asteroids outside the family, in 
the inner asteroid belt (see 
Table~\ref{table: V-ast}). In this work we 
investigate the possibility that these objects are former family 
members that migrated to their current positions via the 
interplay of Yarkovsky effect and nonlinear secular resonances.

The main dynamical feature of 956 Elisa and 809 Lundia is that 
they are currently inside the $2(g-g6)+s-s6$ ($z_2$ by Milani and 
Kne\v{z}evi\'c, 1993) secular resonance.
Our investigations show that members of the Vesta dynamical family 
may drift in three-body and weak secular resonances until 
they are captured in the strong $z_{2}$ secular resonance.  Only 
asteroids with diameters larger than 16 km can remain in one of 
the three-body or secular resonances long enough to reach the 
region of the $z_2$ resonance.  This two-step mechanism of capture into 
the $z_2$ resonance could explain:  i) the current resonant orbits of 
956 Elisa and 809 Lundia, ii) why their size is significantly 
larger than that of the typical member of the Vesta family, and iii) 
provide a lower limit on the Vesta family age.  We 
believe that other V-type asteroids could have followed the 
same path, and could currently be inside the $z_2$ resonance.  

In an incoming article of this series we will investigate the role that
other mechanisms of dynamical mobility, such as close encounters with 
massive asteroids, may have played in causing 
the current orbital distribution of the remaining 20 other V-type 
asteroids.

   }
   \maketitle
%

\section{Introduction}

Asteroid families are thought to be formed as the results of collisional
events in the main belt.  After the break-up or cratering event 
on the parent body, the fragments are ejected and form an cluster 
identifiable in proper element space.  Unlike instantaneous orbital elements, 
that  respond to short-period perturbations, proper elements  
remain nearly constant in time for conservative systems 
(Lemaitre 1993).  However, when chaotic diffusion or non gravitational 
forces are considered, proper elements may change significantly.
Since the family formation, some of the members may 
therefore have drifted in proper element space, and could no 
longer be recognizable as part of the family.  By {\em dynamical family} 
we define the family as identified on the basis of the {\em current} 
proper orbital elements.  The problem of identifying possible past members
of asteroid families has been the subject of many recent studies (Bottke 
{\em et al} 2003).  In this paper we will concentrate on the case of 
the Vesta family.

\begin{figure}

  \centering
  \begin{minipage}[c]{0.5\textwidth}
    \centering \includegraphics[width=2.5in]{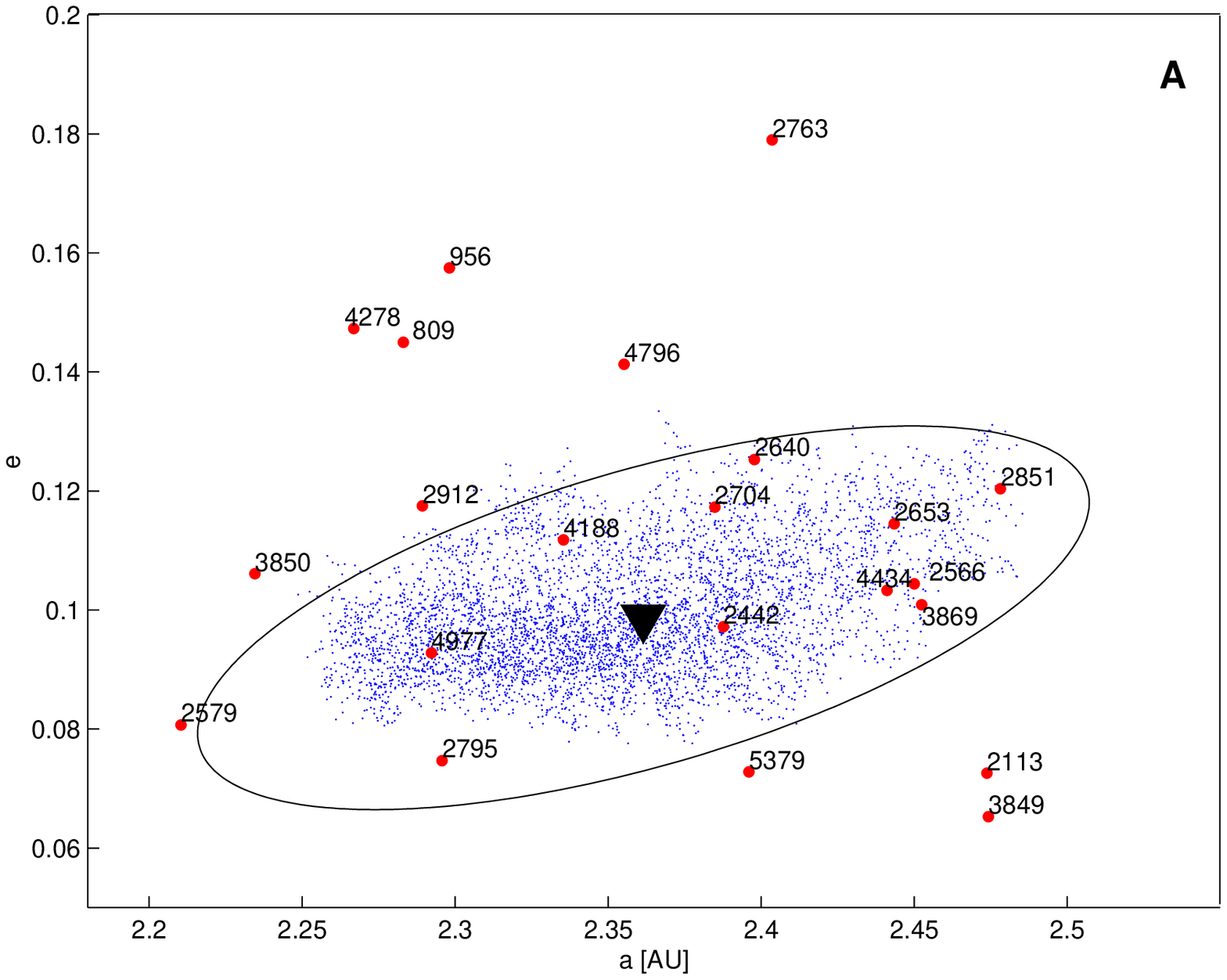}
  \end{minipage}%
  \begin{minipage}[c]{0.5\textwidth}
    \centering \includegraphics[width=2.5in]{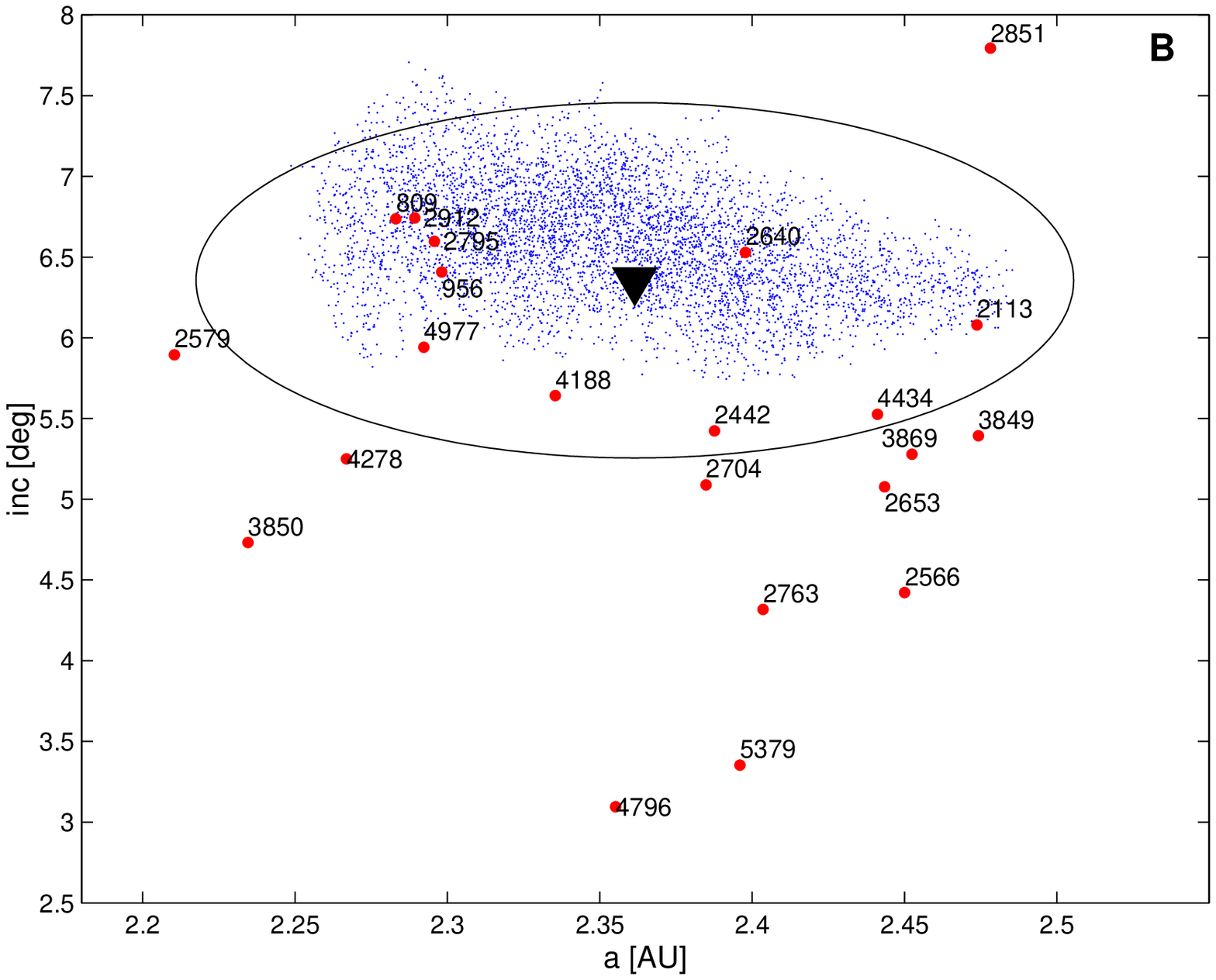}
  \end{minipage}

  \caption{Location in proper element space ({\em a-e}, 
Fig.~\ref{fig: ves_aei}a, {\em a-i}, Fig.~\ref{fig: ves_aei}b) of the 
5112 asteroids (small dots) members of the Vesta family (Moth\`{e}-Diniz
{\em et al.} 2005).  The black triangle shows the 
location of 4 Vesta itself, while the ellipse displays the 
600 m/s level of maximum ejection velocity.
The other dots show the locations of the 22 V-type asteroids that are not 
members of the family.}
\label{fig: ves_aei}
\end{figure}

The Vesta family, one of the largest in the inner asteroid belt, 
is believed to have originated in a cratering event that excavated 
a basin on the surface of 4 Vesta.  According to Marzari {\em et al.} 
(1996) this cratering event occurred $\simeq$~1~Gyr ago (the 
age is poorly constrained).  
Using the hierarchical clustering method (Zappal\`a {\em et al.} 
1990), and a cutoff of 62 m/s, Moth\`{e}-Diniz {\em et al.} 2005 
estimated that the Vesta family presently accounts for about 5000 
members, mostly with diameters of less than 4 km    The typical ejection 
velocities with respect to 4 Vesta necessary to reach the edges 
of this family are of the order of 600 m/s (see Fig.~\ref{fig: ves_aei}).

Most of the members of the Vesta dynamical family (including 4 Vesta itself) 
present a V-type spectrum, which is characterized by a 
moderately steep red slope shortwards of 
0.7 $\mu$m and a deep absorption band longwards of 0.75 $\mu$m.
This kind of spectra is associated with a basaltic surface 
composition (McCord 1970, Bus 2002, Duffard {\em et al.} 2004).
Until the discovery of 1459 Magnya (Lazzaro 2000) 
and of several V-type NEAs (Bus and Binzel 
2002 and references within), the only known 
V-type asteroids were members of the Vesta family.
Recently Florczak {\em et al.} 2002 discovered two inner belt 
asteroids located well
outside the edges of the family (809 Lundia and 956 Elisa) that show 
a V-type spectrum.  Today 22 V-type asteroids that are not thought 
to be members of the dynamical 
family are known in the inner belt (Fig.~\ref{fig: ves_aei}, 
Table~\ref{table: V-ast}).   Most of these asteroids have orbits 
corresponding to ejection velocities with respect to 4 Vesta 
larger than 1 km/s, the maximum possible ejection velocity 
expected to be produced in the cratering event.

The V-type asteroids outside the family can
either be former members that dynamically migrated to their
current orbital positions after family formation, or fragments
of primordial large V-type asteroids other than 4 Vesta.  This second 
scenario opens interesting perspectives about the primordial main 
belt.  However, before considering it, we should be certain 
that all possible migration mechanisms from 
the Vesta family have been dismissed as improbable or impossible.

Among the possible mechanisms of dynamical migration there is evolution 
in secular resonances.  A secular resonance occurs when the frequency 
of variation of the longitude of pericenter $\varpi$ (g), or longitude
of node $\Omega$ (s), becomes nearly equal to an eigenfrequency (or a 
combination of eigenfrequencies) of the system of coupled planetary orbits.
The main linear secular resonances (of order 2), such as $g-g_5$, $g-g_6$, and 
$s-s_6$ , have been recognized in the asteroid belt for a long time
(Williams and Faulkner 1981).  But, apart for the works of 
Milani and Kne\v{z}evi\`{c} (1992, 1993), much less attention has 
been paid to the study of secular resonances of higher order 
(nonlinear secular resonances).

It has been long believed that the effect of nonlinear secular resonances 
on the dynamical mobility of asteroids could be neglected when compared with 
chaotic diffusion in two- and three-body mean-motion 
resonances.  Previous studies 
(Michtchenko {\em et al.} 2002, Lazzaro {\em et al.} 2003) showed 
that chaotic diffusion in nonlinear secular resonances was simply too 
slow to produce significant dispersion of simulated members of 
asteroid families, even over timescales of 2 Gyr or more.

Evolution in nonlinear secular resonances can be much faster 
when the Yarkovsky effect is taken into consideration. 
The Yarkovsky effect is a thermal radiation force that 
causes objects to undergo semimajor axis drift as a function of their 
spin, orbital, and material properties (Bottke {\em et al.} 2001).  It 
is essentially caused by the thermal re-emission of light by the asteroids, 
and is present in diurnal (which can be dissipative, i.e., 
move asteroids towards smaller semimajor axis for retrograde rotators, 
and anti-dissipative, for prograde rotators) and seasonal (always dissipative)
versions.

Bottke {\em et al.} 2001 showed that the unusual shape in the proper
{\em a-e} space of the Koronis family could be explained by 
evolution of family members in the $g+2g_5-3g_6$ 
nonlinear secular resonances via the Yarkovsky effect.
Vokrouhlick\`{y} {\em et al.} 2002 suggested that evolution inside
the $g-g_6+s-s_6$ (named $z_1$ following Milani and Kne\v{z}evi\`{c} 1993) 
secular resonance could explain the presence of K-type
asteroids associated with the Eos family outside its boundaries.  
Our unpublished results for the 
Maria family also suggest the importance of diffusion in nonlinear 
secular resonances when the Yarkovsky effect is considered.

Since several V-type asteroids outside the Vesta family are in a region where  
previous studies of Milani and Kne\v{z}evi\`{c} (1993) have identified
several nonlinear secular resonances able to cause significant long-term 
variations in eccentricity and inclination, we
believe that evolution in these secular resonances 
combined with Yarkovsky effect could have played a role in their 
dynamical evolution.

In this work we try to understand how the interplay of the two effects
could have caused the 
diffusion of V-type asteroids outside the boundaries of the Vesta family.
This work is so divided:  in the second section we consider the general 
dynamics in the region of the Vesta family, in section 3 we 
concentrate on the case of 
three asteroids in the $2(g-g_{6})+s-s_6$ (named $z_2$ after Milani and 
Kne\v{z}evi\`{c}, 1993)\footnote{A $z_k$ resonance is a resonant combination 
of the form $k(g-g_6)+s-s_6$} secular resonance
and their dynamical evolution when the Yarkovsky effect is considered.  
In section 4 we study mechanisms of migration to the $z_2$ secular resonance
from the Vesta family.  In section 5 we present our conclusions and we 
try to identify other asteroids that could have migrated from the Vesta 
family. 


\section{Secular resonances in the Vesta family region}

\begin{figure}
   \centering
   \includegraphics[width=0.75\textwidth]{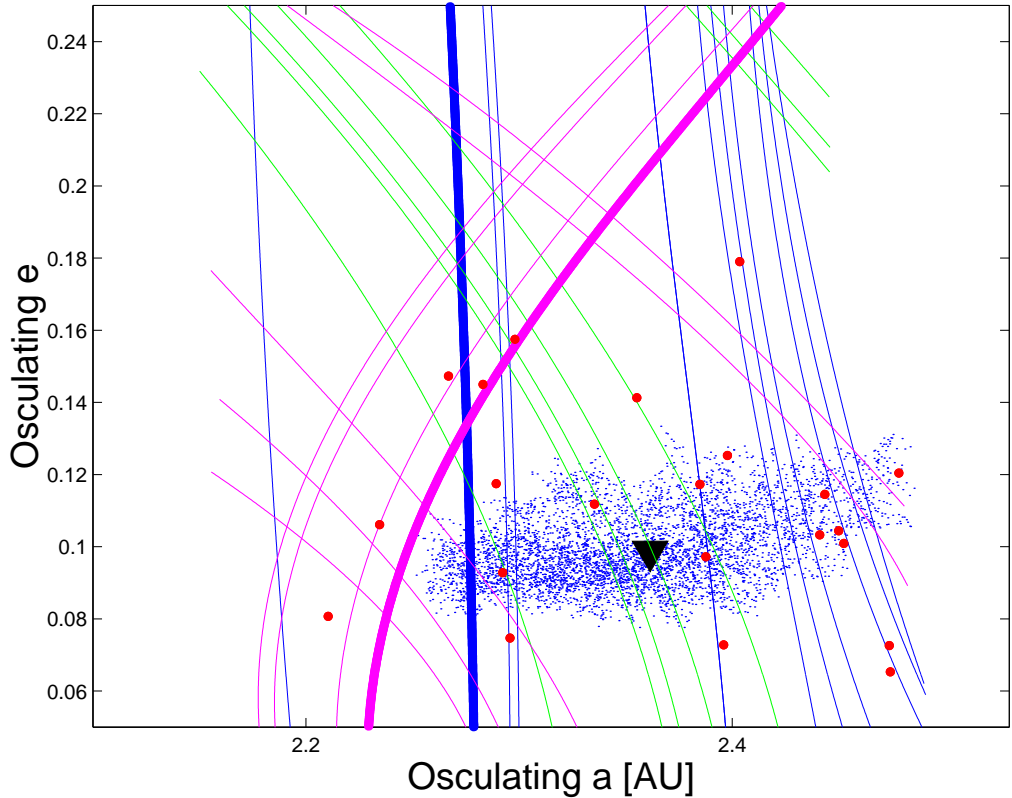}
      \caption{Location of the nonlinear secular resonances in the 4 Vesta 
region, computed for the inclination of 4 Vesta ($i = 6.357^{\circ}$), 
in the osculating $a-e$ plane, determined with the 
Spectral Analysis Method (SAM hereafter, Michtchenko {\em et al.} 2002).  
In blue we show the resonances of perihelia, in green those of nodes, and 
in magenta those of both perihelia and nodes. For reference we also 
show the locations in proper element space of 4 Vesta (triangle) 
the Vesta family members (small dots) and of the V-type 
asteroids outside the family (full dots).  The thick lines display the 
location of the $z_2$ (in blue) and ${\gamma}_2$ (in magenta) 
resonances.  For the identification 
of the other nonlinear secular resonances see 
Table~\ref{table: sec-per-nod}).
              }
         \label{fig: sec-res-tat}
   \end{figure}

The region occupied by the Vesta family is crossed by a very 
rich and complex web of 
secular resonances.  To identify the main nonlinear secular resonances
in the region we integrated a grid of 3131 particles \footnote{101 initial 
values of $a$ in the range 2.1-2.5, and 31 initial values of $e$, 
between 0-0.3, with the other elements the same as Vesta's.}.  
We computed the $g$ and $s$ frequencies of the particles and determined 
all the secular resonances up to order six.  Table~\ref{table: sec-per-nod} 
reports a list of the principal nonlinear secular resonances of perihelion, 
node, and both perihelion and node, respectively, that we found with this
method, with the name given in this paper.  
Fig.~\ref{fig: sec-res-tat} 
displays the locations of the non linear secular resonances in the 
osculating {\em a-e} plane determined with the 
Spectral Analysis Method (SAM hereafter, Michtchenko {\em et al.} 2002).  
For reference, we also show the location 
(in proper elements space) of Vesta, the family members, and the 28  
V-type asteroids outside the family currently known in the inner belt.

Several V-type asteroids seem to be in or very close to secular resonances in
Fig.~\ref{fig: sec-res-tat}, and the reader might be mislead to believe 
that most of them are currently being affected by such resonances.  
This is actually far from being true.  In Fig.~\ref{fig: sec-res-tat} 
we show the location of the secular resonance in the osculating $a-e$ plane, 
while the V-type asteroids (shown only for illustrative purposes) are in 
proper element space.  The actual positions of the resonances 
are of course different in proper element space.  Another problem with 
Fig.~\ref{fig: sec-res-tat} is that secular resonances can be completely 
identified only in the three-dimensional proper $(a,e,i)$ space.
The location of secular resonance in 
the $a-e$ plane depends on the inclination for which it was computed 
(Milani and Kne\v{z}evi\`{c} 1993).  Since each V-type asteroid has 
a different inclination, the location of the web of secular resonance
in the $a-e$ plane is unique for each asteroid.

To understand whether any of the asteroid is involved (or close to) 
a secular resonance, we integrated the 22 V-type 
asteroids outside the Vesta family evolving 
under the influence of the planets, 
from Venus to Neptune over six million years (the mass of Mercury was added 
to that of the Sun; hereafter this will be our standard Solar 
System model, unless otherwise specified), and computed 
the resonant argument of each nonlinear secular resonance for each asteroid.
Table~\ref{table: V-ast} reports the 22 V-type
asteroids names and numbers, proper {\em a,e, i},
absolute magnitude H, and if they are in or close to a 
nonlinear secular resonance (in this 
case we report the approximate circulation period of the resonant angle).

Among all the detected secular resonances,
the $z_2$ and the $2g-2g_6+g_5-g_4$ (${\gamma}_2$ hereafter)
seem to be the most interesting, since they are the ones 
that are affecting the largest number of asteroids.
Two asteroids are currently inside the $z_2$ secular resonance 
(809 Lundia and 956 Elisa), and three asteroids are very close
to the ${\gamma}_2$ resonance (4977 Rauthgundis, 956 Elisa, 
and 2795 Lepage).  As pointed out by 
Milani and Kne\v{z}evi\`{c} (1993), the $z_2$ resonance, 
being a secondary mode of the $g-g_6$ resonance, is one of the strongest 
nonlinear secular resonance in the inner belt.  The $2g-2g_6+g_5-g_4$
is also expected to be a strong resonance, since it contains the 
combination $2{g_{6}}-g_{5}$, which is the leading critical term of the 
secular planetary theory (Nobili {\em et al.}, 1989).

\begin{figure}

  \centering
    \centering \includegraphics[width=0.75\textwidth]{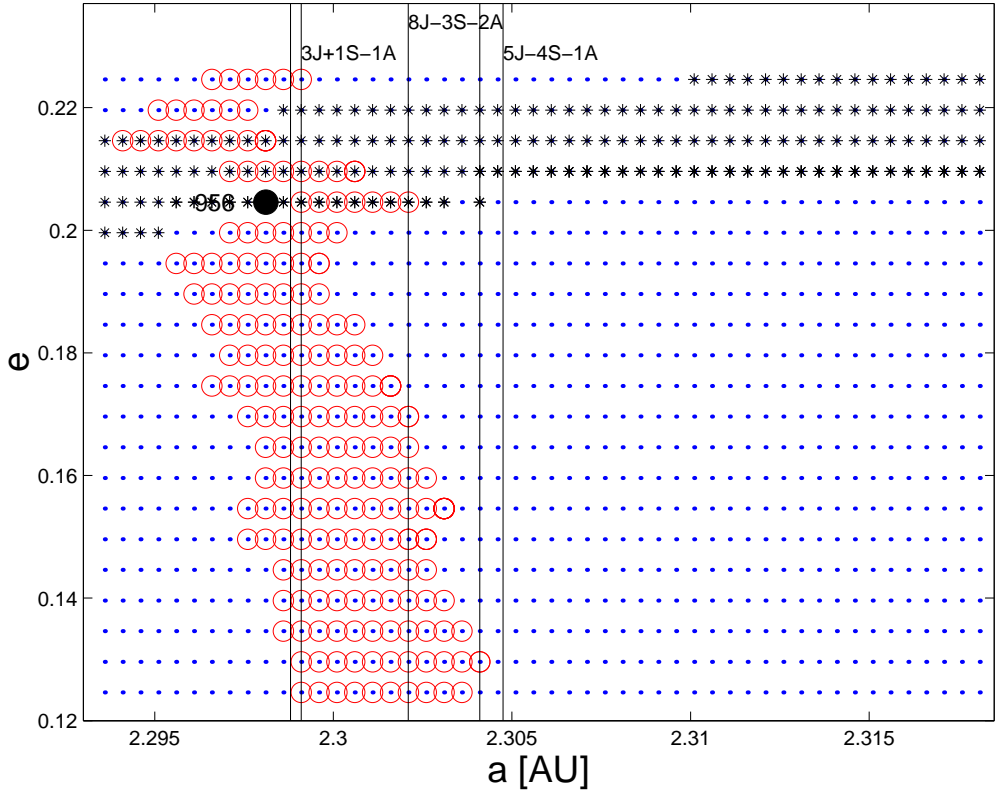}

     \caption{The plot shows the location of the $z_2$ ($2(g-g_6)+s-s_6$)  
       secular resonance (black asterisks), of the ${\gamma}_2$ 
       ($2g-2g_6+g_5-g_4$) (red circles), and of the 3J+1S-1A, 8J-3S-2A, 
       and 5J-4S-1A three-body resonances (vertical lines), as found 
       by plotting the resonant arguments of the test particles.   The 
       black full circle displays the actual position of 956 Elisa, and 
       small dots show all the initial conditions used.
       The 3J+1S-1A resonance lies in the band from 2.29880 to 
       2.29910 AU (as compared with the analytical result of 
       Nesvorn\`{y} and Morbidelli 1998 of 2.2994 AU).}
     \label{fig: Elisa-dyn}
\end{figure}

In view of this, it would be necessary to determine the location in 
the osculating $a-e$ space of these two resonances, in particular in the 
region of 956 Elisa, which is the asteroid currently inside the $z_2$ 
and very close to the ${\gamma}_2$ resonances.  This is not as 
simple a task as it may seems. Contrary 
to the case of the most important linear secular resonances ${\nu}_6$, 
${\nu}_5$, and ${\nu}_{16}$, whose dynamics has been studied in great detail 
by Morbidelli and Henrard (1991a, b), no analytical model has ever
been made for nonlinear secular resonances.  To study orbits
inside these nonlinear resonances we cannot  
use the approach of Migliorini {\em et al.} 1997 for 
the ${\nu}_6$ resonance, where a grid of test 
particles in initial osculating $a-e$ was created and the 
inclination was determined by the requirement of being at the exact 
resonance.  We need to find an alternative procedure
to map the new resonances.

To determine the location of a nonlinear 
secular resonance there are essentially three 
methods:  i) use the second order and fourth-degree secular 
perturbation theory of Milani and 
Kne\v{z}evi\`{c} (1990) to find the $g$ and $s$ precession rates of the
longitudes of perihelion and node of asteroids, and, therefore, the 
locations where such values are close to a resonance in proper element 
space; ii) compute the frequencies numerically{\bf ,} analyzing the 
equinoctial elements $(e\cos{\varpi}, 
e\sin{\varpi})$, $(\sin{(I/2)}cos{\Omega}, \sin{(I/2)}sin{\Omega})$ of test 
particles to obtain $g$ and $s$ with the frequency modified Fourier 
transform (FMFT, hereafter) 
of \v{S}idlichovsk\`{y} and Nesvorn\`{y} (1997) or with SAM and, 
therefore, the location of the resonances in osculating space; or iii) plot 
the resonant argument of the secular resonances and define the boundary 
as the place where we pass from circulation to libration.

The Milani and Kne\v{z}evi\`{c} approach, while very useful to obtain 
a qualitative understanding of secular resonances' dynamics, has 
some drawbacks when you are interested in precisely determining 
the location of secular resonances in a numerical 
simulation.  First, the Milani and Kne\v{z}evi\`{c} method 
does not include the effect of the perturbations of the terrestrial 
planets, that are important for the dynamics in the Vesta
family region. Also, it is an analytical model 
based on a linear theory, so the values of the asteroid and 
planetary frequencies (and therefore the locations of the 
secular resonances) are usually different from 
those computed with a numerical simulation.  Since the actual 
location of the secular resonances is essentially unique to
each different simulation (for example, neglecting the 
indirect effects of the 
perturbations of the terrestrial planets on the jovian ones 
may change the planets frequencies by up to 0.05 $''$/yr), 
in this work we prefer to use a numerical technique.  

Of the two numerical methods, determining the resonances by computing 
the frequencies of the particles presents two disadvantages.
To apply this method we need to numerically 
determine the values of the proper frequencies.
Small errors in the computation of $g_4$, for example, 
may lead to an erroneous determination of 
the $2g-2g_6+g5-g_4~({\gamma}_2)$ 
resonance.  Moreover, this method does not give any information on the 
resonance width.  For these reasons, we decided to use only the last 
method, i.e., plotting the resonant argument.

To map the $z_2$ and ${\gamma}_2$ resonances we 
generated a grid of 1250 test particles, with 50 initial values of
$a$ values from 2.29360 to 2.31810 AU, and 25 initial values of $e$,
ranging from 0.12 to 0.21, and all the other angles 
the same as 956 Elisa.  
We integrated them over 6 Myr, enough to sample
at least one libration period of both resonant arguments, plotted the 
resonant argument of each test particle, and determined where 
it passed from circulation to libration.  Fig.~\ref{fig: Elisa-dyn} shows 
the locations of the two resonances in the initial osculating $a-e$ plane.

So far, we have determined the sections of the two resonances in 
the initial osculating $a-e$ plane.
To completely study the resonances, we should have done two other series of 
simulations, so as to obtain the resonances' sections 
in the $a-i$ and $e-i$ planes, and this for each of the 28 V-type 
asteroids outside the Vesta family!  Rather than following 
this computationally expensive approach, we decided to study the 
dynamics of asteroids inside the $z_2$ resonance by integrating 
clones of the three resonant asteroids under the influence of 
the Yarkovsky force.  This approach
not only could tell us if the Yarkovsky force may take
the asteroids out of the $z_2$ resonance, but also give us precious
hints on the asteroids' origins. We will discuss the results of 
this approach in more detail in the next section.


\section{Evolution in the $z_2$ secular resonance under the Yarkovsky effect}

In the previous section we identified a web of secular resonances in the 
region of the Vesta family, and determined which asteroids are in or close to 
a nonlinear secular resonance.  It turns out that only two asteroids are 
currently in a nonlinear secular resonance: 809 Lundia and 956 Elisa, 
which are inside the $z_2$ resonance (see 
Table~\ref{table: V-ast-sg4}).

In this work we are interested in the origin of V-type asteroids 
outside the Vesta dynamical family.  Usually, studies on 
Yarkovsky diffusion of asteroid families use the approach of numerically
integrating test particles simulating the orbital initial 
conditions of the family members immediately after the family formation.
These test particles are evolved under the influence of the Yarkovsky 
effect and allowed to interact with the local web of resonances 
(mean-motion and secular).  The analysis of their evolution provides 
hints on their possible paths outside the dynamical 
families (Bottke {\em et al.} 2001).

\begin{figure}

  \centering
  \begin{minipage}[c]{0.5\textwidth}
    \centering \includegraphics[width=2.5in]{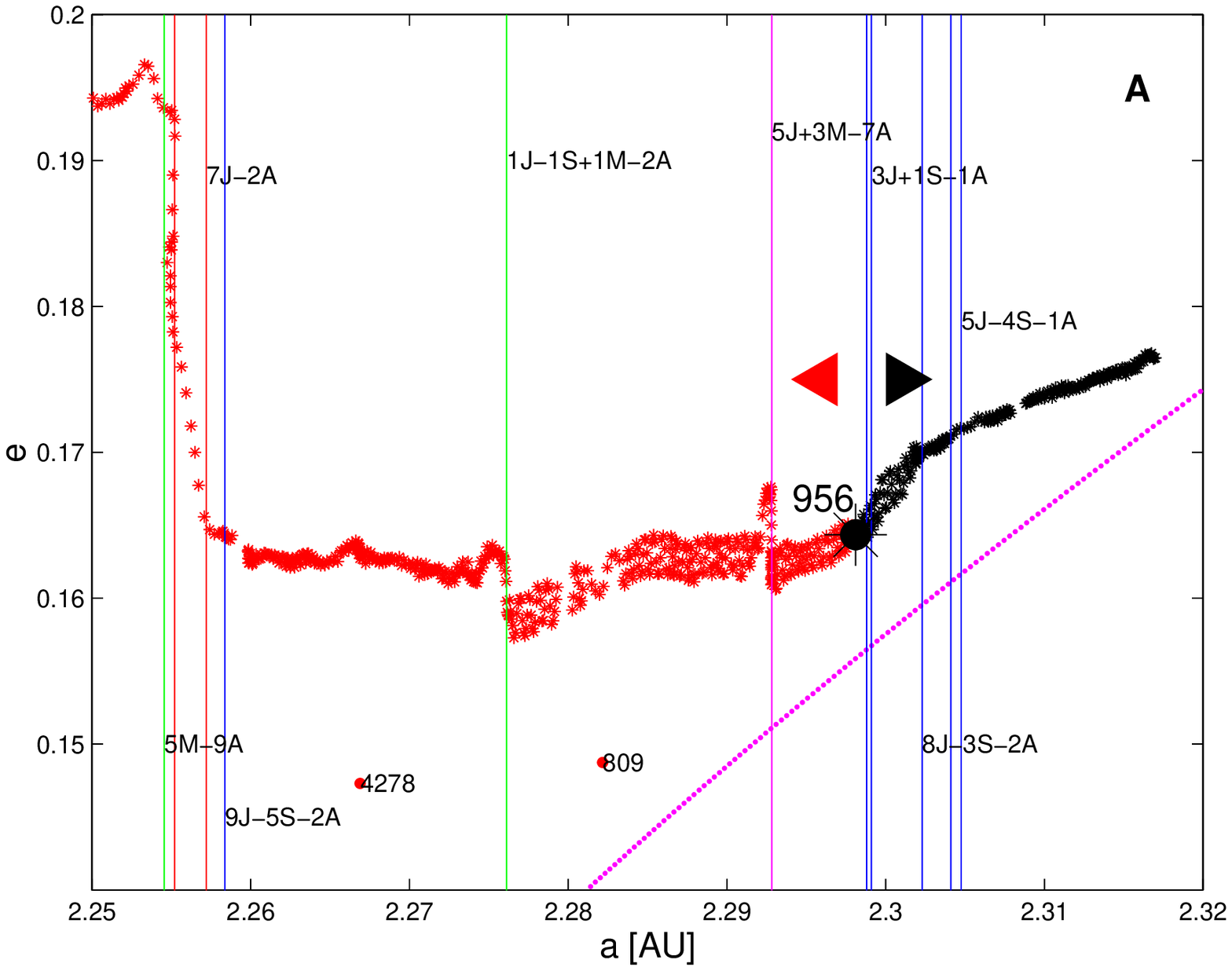}
  \end{minipage}%
  \begin{minipage}[c]{0.5\textwidth}
    \centering \includegraphics[width=2.5in]{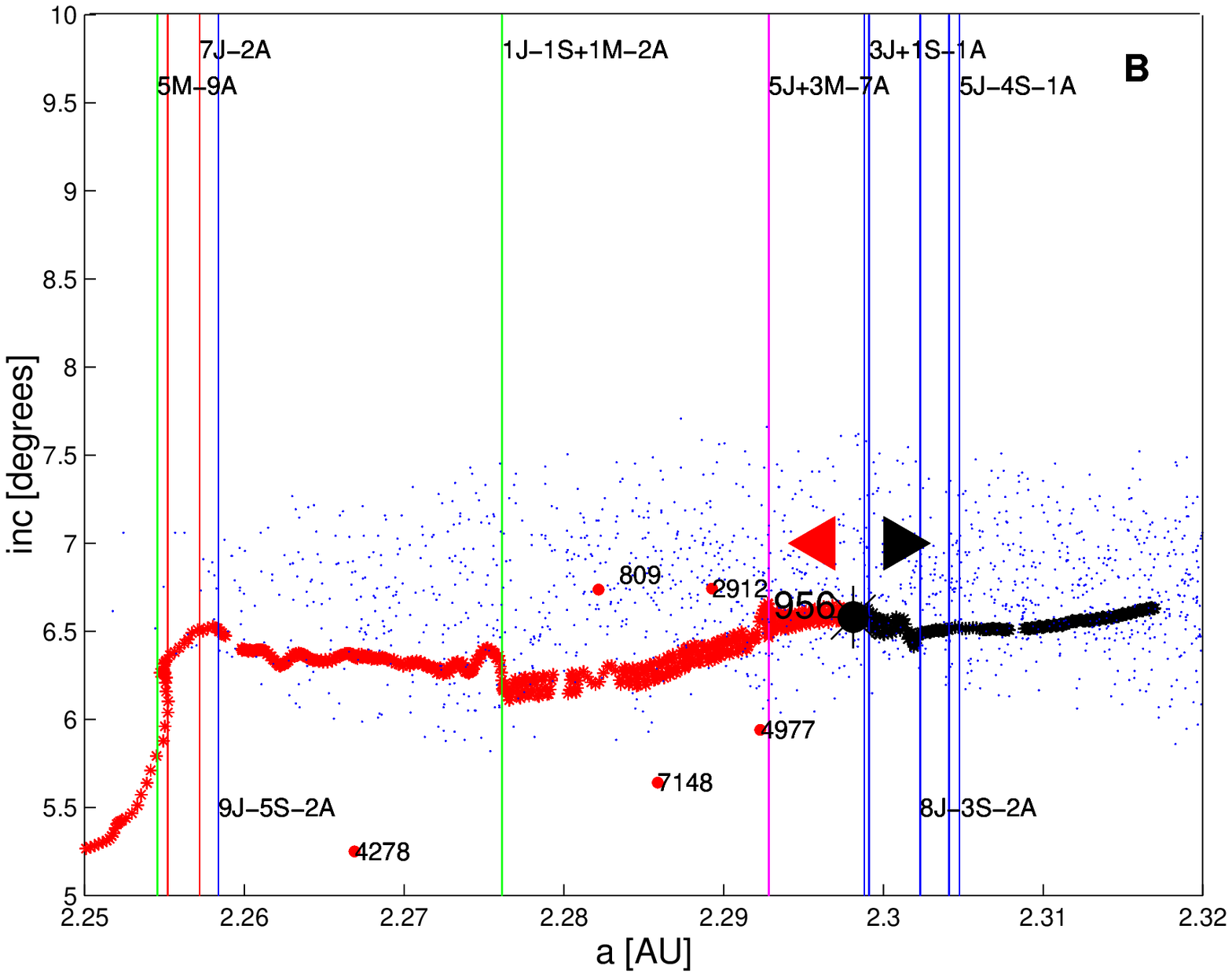}
  \end{minipage}

\caption{The averaged {\em a-e} and {\em a-i} 
evolution of two 100 m clones of 956 Elisa.  In black we show the evolution 
of the prograde clone, in red of the retrograde one.  The arrows show the 
two directions of evolution in {\em a}, starting from 956 Elisa (black 
full dot).  Small dots show the orbital location of 
members of the Vesta dynamical family.  The magenta line in 
Fig.~\ref{fig: clones_aei_elisa}a gives the 
location of the $z_2$ secular 
resonance, computed for the inclination of Vesta (see 
Fig.~\ref{fig: sec-res-tat}) and is given for reference. 
Note how the clone of Elisa crosses the three-body resonances 
(3J+1S-1A and 5J-4S-1A, where hereafter J stands for Jupiter, 
S for Saturn, M for Mars, U for Uranus, and A for asteroid), while 
remaining inside the $z_2$ resonance (its path in the
$a-e$ plane is inclined, while it should be a horizontal line
if the particle were freely drifting due to Yarkovsky effect).  
The clone leaves the resonance 
only when interacting with the 1J-1S+1M-2A four-body resonance.  It then 
freely drifts due to the Yarkovsky effect until it 
reaches the 7J-2A resonance where it is temporary captured.}
\label{fig: clones_aei_elisa}

\end{figure}

This approach has both merits and drawbacks.  We have seen that 
nonlinear secular resonances can be very narrow in osculating elements space.  
Therefore, by integrating clones with random initial osculating 
elements, their interaction with a specific secular resonance 
might be overlooked.  Moreover, in this case we are in a rather 
different situation than the typical one.  We not only know where 
possible former family members are presently located, but we also know
that some of these bodies are currently interacting with a 
specific nonlinear secular resonance. 

In view of these considerations, we adopted a different approach.
We chose to integrate clones of the two asteroids currently 
inside the $z_2$ resonance with the Yarkovsky effect. 
Since the Yarkovsky effect depends on asteroid size (smaller 
objects drift faster), we considered clones 
with the same initial orbital parameters of the original asteroids,
but with different diameters.
From their dynamical evolution, we try to infer
the possible paths that might have brought them 
to their current orbital position with respect to the Vesta family.  
This approach not only can give hints at the past 
dynamical history of the two asteroids, but can tell us
for which sizes the Yarkovsky 
drift can drive the bodies out of the $z_2$ resonance. 

To simulate the Yarkovsky effect we used SWIFT-RMVSY, the version 
of SWIFT modified by Bro\v{z} 1999 to account for both versions of the
Yarkovsky effect.  We created 10 clones each with the same orbital elements 
of 809 Lundia, 956 Elisa, and 
6406 1992 MJ, with radii of 100m, 500 m, 1 km,  2 km, 3 km, 4 km, 
5 km, 6 km, 7 km, and 8 km (the maximum possible size of the two asteroids)
and integrated them with a step size of 20 days.  We referred all 
elements with respect to the invariable plane of the Solar System, and 
used the following thermal 
parameters: ${\rho}_{bulk}$ = 3500 $kg/m^3$, ${\rho}_{surface}$ = 1500 
$kg/m^3$, $K = 2.65 W/m^3/K$, $C = 680$~$J/Kg/K$, A =0.1, and $\epsilon$ = 
0.9 (Farinella et al. 1998).  

All asteroids were either prograde rotators with 
$0^{\circ}$ obliquity, (so that the diurnal Yarkovsky effect was 
anti-dissipative, and the semimajor axis evolved towards larger values), 
or retrograde rotators (evolution towards smaller {\em a}), with 
$180^{\circ}$ obliquity.  The period of rotation was assumed 
proportional to the inverse of the radius as in Farinella {\em et al.} 
1998.  Since our goal was to study diffusion in the $z_2$ secular 
resonance, we did not consider reorientations of spin axis via collisions or 
YORP (\v{C}apek and Vokrouhlick\`{y} 2004).  Our 
simulated asteroids have constant obliquity 
and rotation periods.  This assumption will not give realistic results for 
the timescale of dynamical evolution of asteroids, but will allow them to more 
efficiently evolve in  the $z_2$ secular resonance.\footnote{We 
should point out
that our assumption of zero obliquities may not 
be that naive.  Studies of \v{C}apek and Vokrouhlick\`{y} 2004 show that 
when thermal conductivity is considered, asteroids do tend to evolve towards 
zero obliquities (and progressively slower or faster spins rates).} 

All asteroids under study are in a region that is roughly 
delimited by two strong mean-motion resonances: the 7J-2A and the 5J-4S-1A
(roughly speaking, $a$ going from 2.255 to 2.304 AU).
In our work we will concentrate only on this region of space.
We integrated the asteroids for 1.5 Gyr, 
a time that was sufficient to allow the 100 m 
clones to cover the region between the two mean-motion 
resonances.  By integrating smaller asteroids than the actual ones inside the 
$z_2$ resonance we could obtain a faster Yarkovsky drift and study 
diffusion in  the resonance 
on timescales much shorter than those required for km-sized objects.

We filtered the osculating orbital elements 
(Carpino {\em et al.} 1987) so as to eliminate all frequencies with periods
smaller than 1333 yrs, and computed synthetic proper elements 
(Kne\v{z}evi\`{c} and Milani 2000) for each $\simeq$ 10 Myr interval
\footnote{The proper semimajor axis was computed by averaging the filtered
elements over 10 Myr.  The synthetic eccentricity and inclination were 
determined analyzing the equinoctial elements ($ecos{\varpi}, esin{\varpi};
sin{\frac{I}{2}}cos{\Omega},sin{\frac{I}{2}}sin{\Omega}$) with FMFT.  
The amplitudes associated with the largest frequencies were the synthetic 
proper $e$ and $i$}.
We also computed averaged elements with a running window of 10 Myr and a shift
of 1 Myr (Nesvorn\`{y} and Morbidelli 1998). While averaged elements are not
constants of the motion like proper elements, they give qualitative 
information on the dynamical evolution of asteroids and are 
easier to compute.   Our results show that apart from a shift of $\simeq0.05$
in $e$ and $\simeq0.2^{\circ}$ in $i$, the qualitative behavior is exactly the 
same either for averaged or proper elements.  For simplicity, we 
will only show averaged elements in our figures.

\begin{figure}
\centering 
\includegraphics[width=0.55\textwidth]{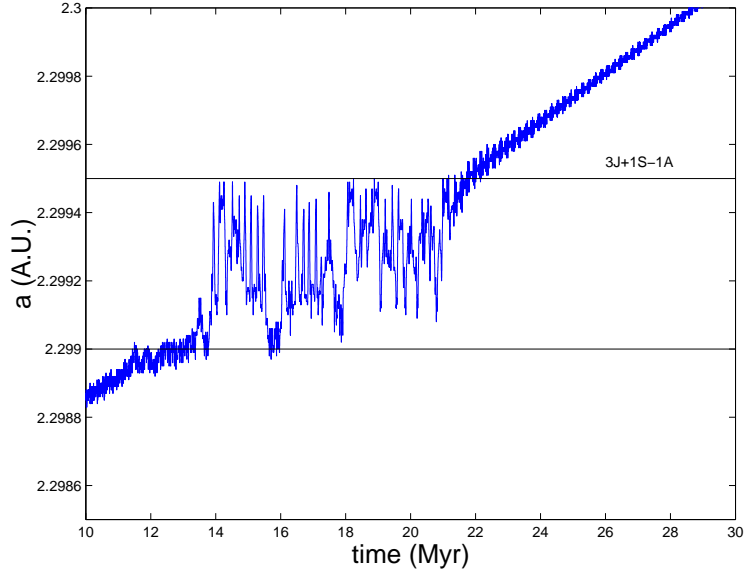}

\caption{Evolution of the osculating semimajor when the 100 m clone of 956 
Elisa crossed the 3J+1S-1A mean motion resonance.}
\label{fig: part11}
\end{figure}

We start by showing the results for 956 Elisa because the dynamics nearby 
is particularly interesting.  As seen in Fig.~\ref{fig: Elisa-dyn}, 956 
Elisa is near two three-body resonances and a nonlinear secular resonance.
In the next section we will show how members of the dynamical family 
could migrate to the $z_2$ by capture in one of these resonances via Yarkovsky 
effect. The simulations of the 100 m clone, however, already offers 
interesting results.  Figs.~\ref{fig: clones_aei_elisa} show the averaged 
{\em a-e} and {\em a-i} evolution of two 100 m clones of Elisa (averaged 
elements computed over a period of 10 Myr).  In red we show the diffusion 
path of the retrograde clone, and in black that of the prograde clone.  The
arrows display the directions of propagation.
Note how the prograde clone 
stays inside the $z_2$ resonance when crossing 
the 3J+1S-1A, the 8J-3S-2A, and the 5J-4S-1A resonances.
Fig.~\ref{fig: part11} shows enlargements of the evolution 
of the semi-major axis of Elisa when crossing the 3J+1S-1A resonances.

\begin{figure}

  \centering
  \begin{minipage}[c]{0.5\textwidth}
    \centering \includegraphics[width=2.5in]{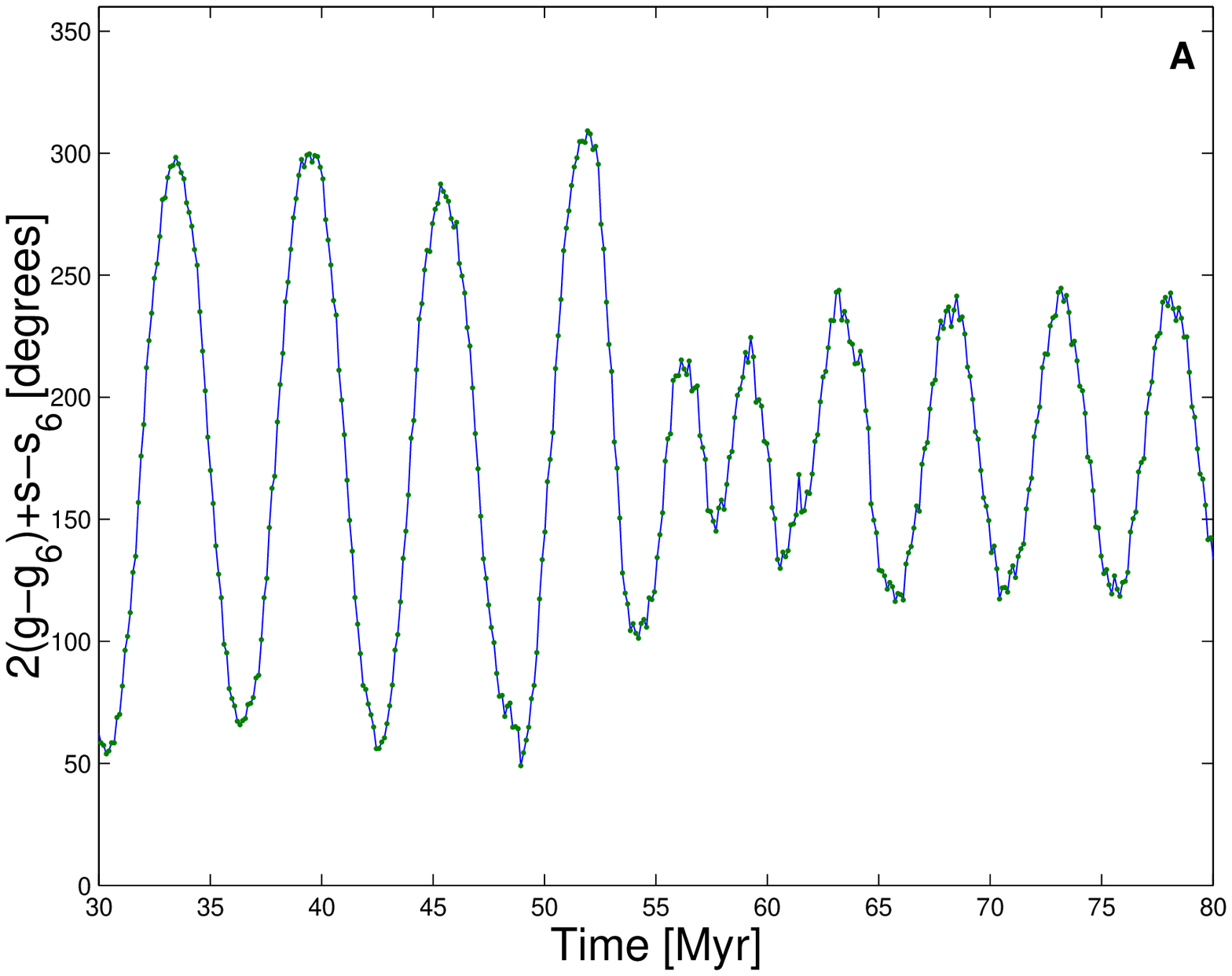}
  \end{minipage}%
  \begin{minipage}[c]{0.5\textwidth}
    \centering \includegraphics[width=2.5in]{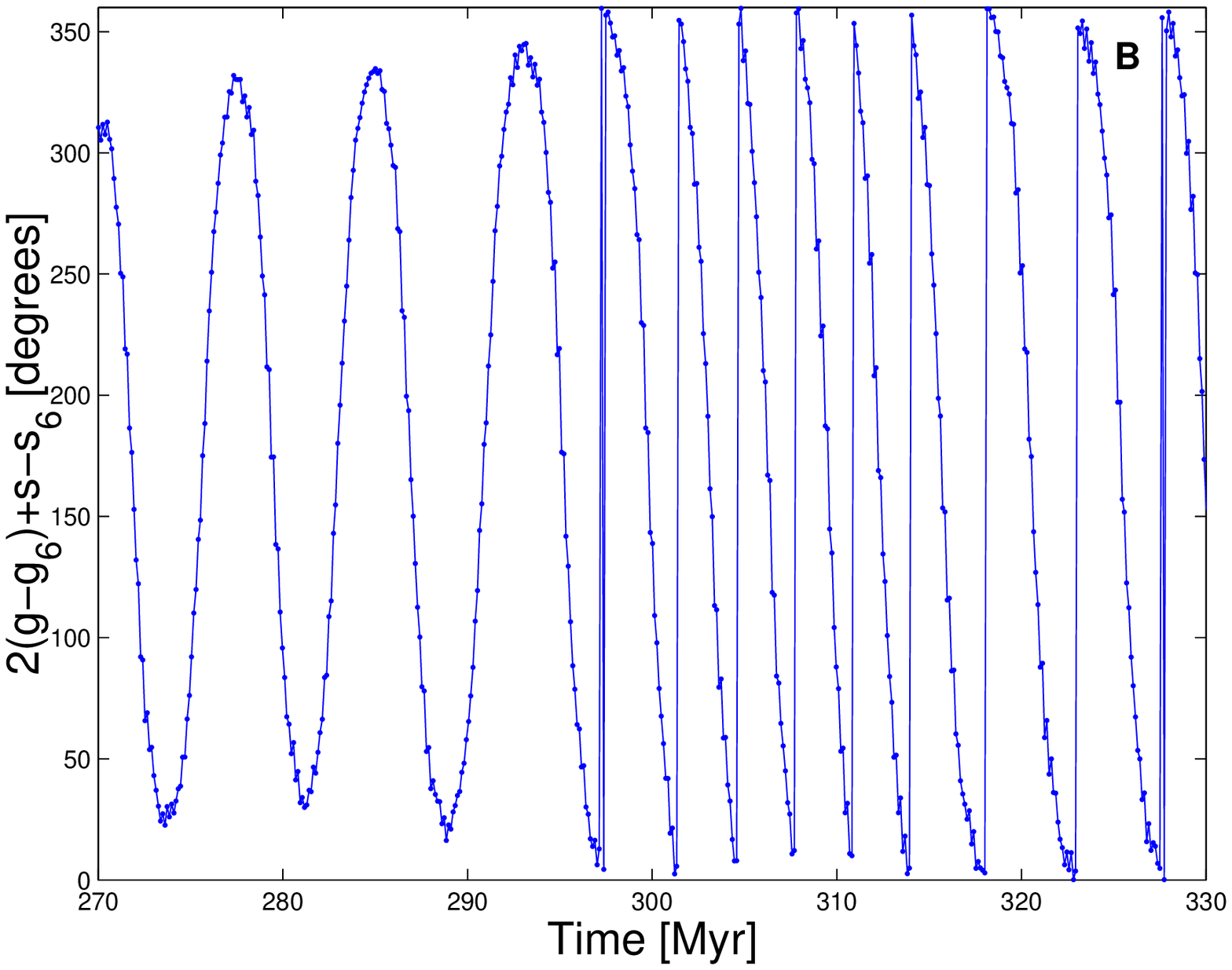}
  \end{minipage}

  \caption{The argument of the $z_2$ resonance for the two  $100~m$ Elisa 
clones evolving under the Yarkovsky effect.  Fig.\ref{fig: part11resarg}a
shows the resonant argument for the clone evolving towards larger $a$, 
when passing through the 8J-3S-2A mean-motion resonance.
Fig.\ref{fig: part11resarg}b shows the argument of the clone 
evolving towards smaller $a$, when crossing the 1J-1S+1M-2A 
four-body resonance (in this case the clone leaves the $z_2$ secular
resonance).  The resonant angles have been processed through a digital
filter in order to remove all frequencies with periods smaller than 
300,000 yr.}
\label{fig: part11resarg}
\end{figure}

The most important fact is that the clone remains inside the $z_2$ 
secular resonance, until it reaches the 1J-1S+1M-2A four-body resonance. 
It then escapes from the secular resonance and drifts freely due 
to the Yarkovsky effect, until it reaches the 7J-2A resonance. 
The particle leaves the secular resonance in the region 
between the 5J-4S-1A and the 7J-2A resonances: this could 
explain the existence of asteroids like 4278 Harvey, that are 
in this region but not in a secular resonance.
After crossing the 8J-3S-2A mean motion resonance, the argument of 
resonance changes libration amplitude, as can be seen in 
Fig.~\ref{fig: part11resarg}a.   

On the right panel of Fig.~\ref{fig: part11resarg}, we 
show the time behavior of the $z_2$ 
resonant argument when the particle crossed the 1J-1S+1M-2A four-body 
resonance.  The fact that the simulated asteroid stayed inside the $z_2$ 
resonance even when this was crossed by powerful three-body resonances
strongly suggest that the $z_2$ could be an efficient 
mechanism to capture objects running away from the Vesta family in 
three-body (such the 5J-4S-1A, or the 3J+1S-1A), or in nonlinear 
secular resonances (such as the ${\gamma}_2$). We will 
further investigate this hypothesis in the next section.

\begin{figure}

  \centering
  \begin{minipage}[c]{0.5\textwidth}
    \centering \includegraphics[width=2.5in]{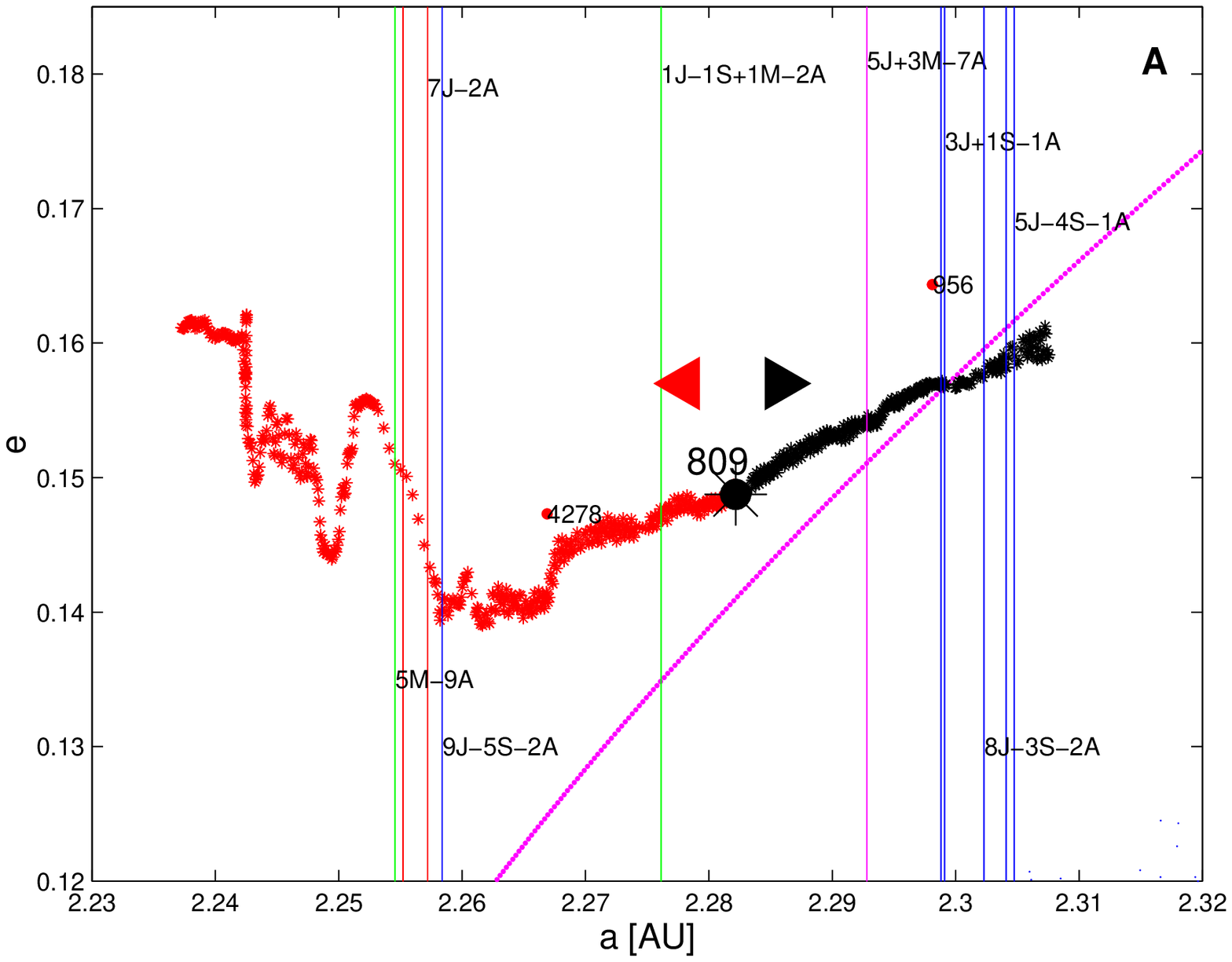}
  \end{minipage}%
  \begin{minipage}[c]{0.5\textwidth}
    \centering \includegraphics[width=2.5in]{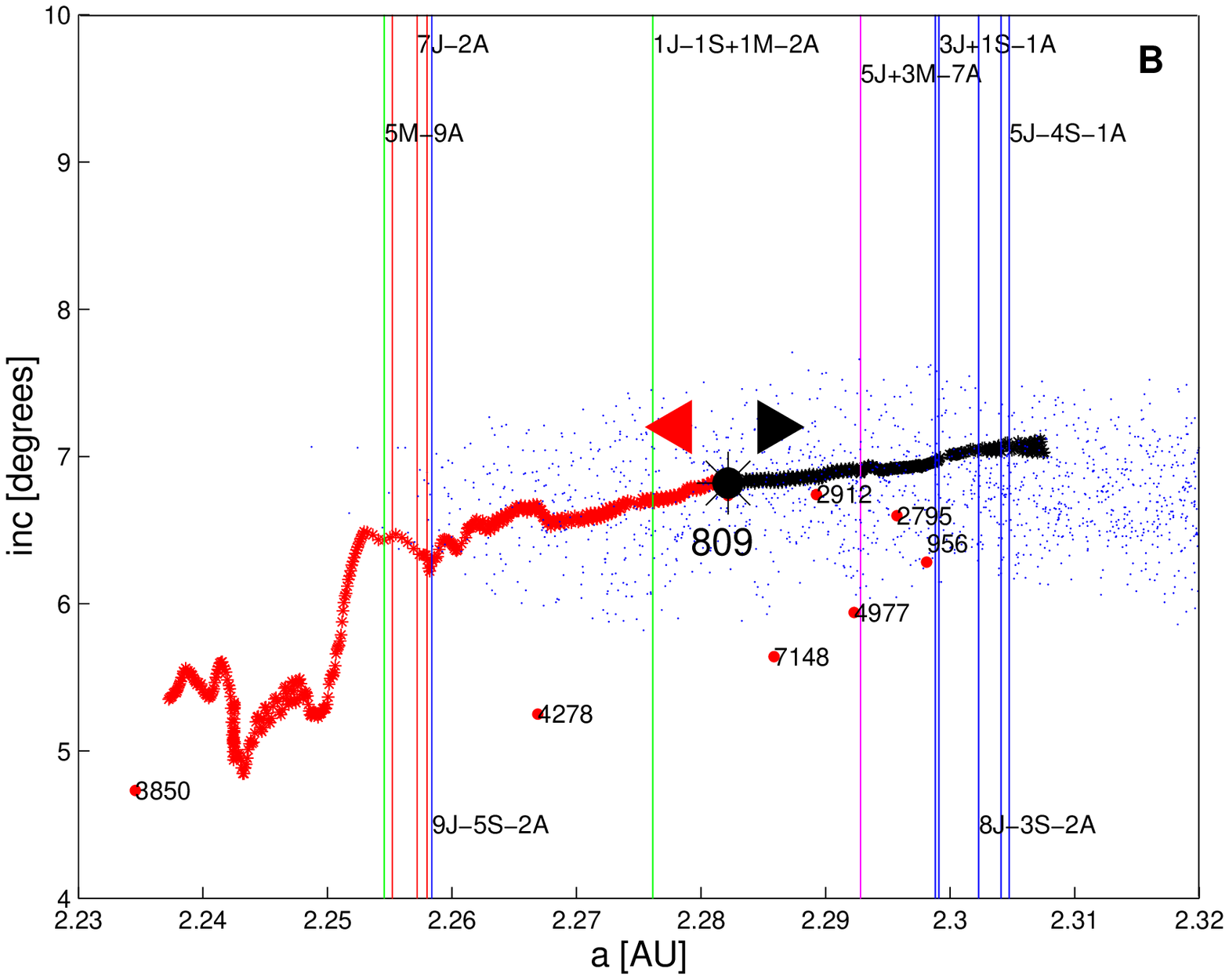}
  \end{minipage}

\caption{The proper {\em a-e} and {\em a-i} 
evolution of two 100 m clones of 809 Lundia.  In black we show the averaged 
elements of the prograde clone and in red those of the retrograde one.  
The arrows show the directions of propagation.  In the region between 
the 7J-2A and the 5J-4S-1A the clone never leaves
the $z_2$ secular resonance.}
\label{fig: clones_aei_lundia}
\end{figure}

The case of the clone of 809 Lundia is easier to interpret.  The clone
essentially followed the secular resonance $z_2$ in the whole region 
between the 
5J-4S-1A and the 7J-2A resonances (Fig.~\ref{fig: clones_aei_lundia}).  In 
the region near the 7J-2A resonance, the $z_2$ resonance interacts with 
other secular resonances.  Once the clone 
reaches the 7J-2A it escaped from the secular resonance.

Finally, we want to see what happens for larger asteroids.  Capture in 
mean-motion resonance is a process that strongly depends on the rate of 
migration of the asteroid which is a function of the asteroid's diameter.
We want to understand for what size capture is possible.  

We start by analyzing the evolution of clones of Lundia.  Since the  
region around Lundia 
is relatively free of resonances, all the clones essentially drift freely 
inside the $z_2$ secular resonance, at a rate that depends on the asteroid's 
size.  We may ask ourselves if diffusion in nonlinear secular 
resonances due to Yarkovsky effect could be as effective as 
diffusion in free space.  In other words, is the Yarkovsky diffusion 
slowed down by the fact that particles are in a secular resonance, or
the diffusion acts at same speed, but with a different geometry 
(following an inclined path in the $a-e-i$ space instead of a horizontal
line of constant $e$ and $i$)?
To answer this question, we computed how much particles inside the 
$z_2$ secular resonance diffused in proper element space, under 
the influence of the other planets from Venus to Neptune, or under the 
influence of the Sun only.    To compute the distance covered in proper 
element space, we used the metric from Zappal\`{a} {\em et al.} 1990:

\begin{equation}
\delta v = na \sqrt{\frac{5 {\delta a}^{2}}{4 a^2}+2 {\delta e}^2
+2 {\delta \sin^2{i}}}
\label{eq: metric}
\end{equation}

\noindent Where $v$ is the distance (in m/s) and $n$ is the 
asteroid's proper motion. We then computed the equivalent 
distance covered in semimajor axis.  
The thermal parameters are the usual for V-type asteroids.  In 
all cases we considered, the distance covered in the secular case is always 
larger than that for the free case.  Changing the metric does not 
change these results by more than 5\%.  This suggests that 
diffusion inside the $z_2$
secular resonance via Yarkovsky effect is at least as fast as diffusion 
via Yarkovsky effect in free space.

\begin{figure}
  \centering
   \includegraphics[width=0.75\textwidth]{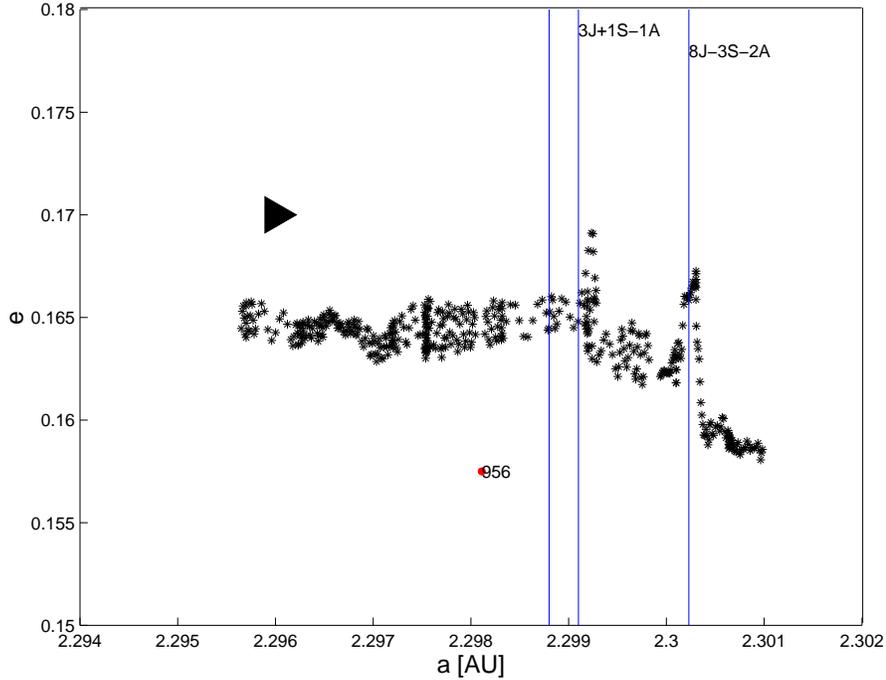}
     \caption{Averaged {\em a-e} evolution of a 4000 m clone of Elisa.  
The particle is temporarily captured into the 3J+1S-1A three-body resonance, 
passes through the 8J-3S-2A resonance, and is finally captured 
into the ${\gamma}_2$ secular resonance.  The black arrow displays
the direction of propagation.}
     \label{fig: Elisa_16}
\end{figure}

The case of Elisa is more interesting.  The clones of Elisa passed through 
the 3J+1S-1A, the 5J-4S-1A, and the ${\gamma}_2$ resonances while 
remaining inside the $z_2$ 
for asteroid radii up to 2000 m.  Capture (and temporary capture) 
into the 3J+1S-1A, 
the 8J-3S-2A and the ${\gamma}_2$ was possible only for larger radii.  
Fig.~\ref{fig: Elisa_16} shows the evolution in the {\em a-e} space of 
a 4000 m clone of Elisa that was finally captured into the 
${\gamma}_2$ secular resonance.
The fact that only particles larger than $\simeq$ 4 km in diameter are captured
into one of the weaker three-body or other secular resonances has interesting 
repercussions on the current population of asteroids inside the $z_2$, 
that will be investigated in more detail in section 4.1.


\section{Mechanisms of migration from the Vesta family to the 
$z_2$ secular resonance}

Simulations with clones of 956 Elisa have shown that asteroids initially 
inside the $z_2$ secular resonance and evolving under the Yarkovsky 
effect tend to stay inside the resonance when crossing 
three-body or other secular resonances and only escape when the resonance
crosses a relatively strong two-body resonance like the 7J-2A. 
This suggests a possible evolution scheme for some Vesta family members:
asteroids are initially captured inside 
three-body resonances like the 5J-4S-1A
or the 3J+1S-1A, or in other secular resonances (such as the 
${\gamma}_2$), where they drift until reaching the zone 
where these resonances cross the $z_2$ resonance.  Then, at 
least a fraction of them is captured into the resonance where 
they slowly drift until reaching the 7J-2A mean motion resonance.
  
In the next subsections we will study in more detail these mechanisms of 
migration, and we will try to address why asteroids currently 
in the $z_2$ resonance have larger diameters than the typical Vesta family 
member.

\subsection{Three-body resonances}

In this section we concentrate on the 5J-4S-1 resonance 
because it is a zero-order 
resonance and previous study (Lazzaro {\em et al.} 2003) have 
shown that chaotic diffusion in this resonance is quite effective 
at increasing asteroid eccentricities to regions close to 
the domain of the $z_2$ secular resonance.  
We believe that our results could qualitatively be extended to the case of 
other three-body resonances.

We simulated members of the Vesta dynamical family by using 
test particles of seven different 
sizes (100, 1000, 4000, 5000, 6000, 7000, and 8000 m), 
just outside the 5J-4S-1A resonance ($a$ = 2.3052 AU), for 
seven values of osculating $e$ (from $e$ = 0.06 to 0.12, 
with a step of 0.1) and the same inclination of Elisa.
Since we are working with retrograde rotators of zero obliquity, our results 
will give an upper limit on the size of objects that can diffuse 
through the resonance.  The actual size limit will depend on the 
cosine of the real obliquity. 

\begin{figure}
  \centering
  \begin{minipage}[c]{0.5\textwidth}
    \centering \includegraphics[width=2.5in]{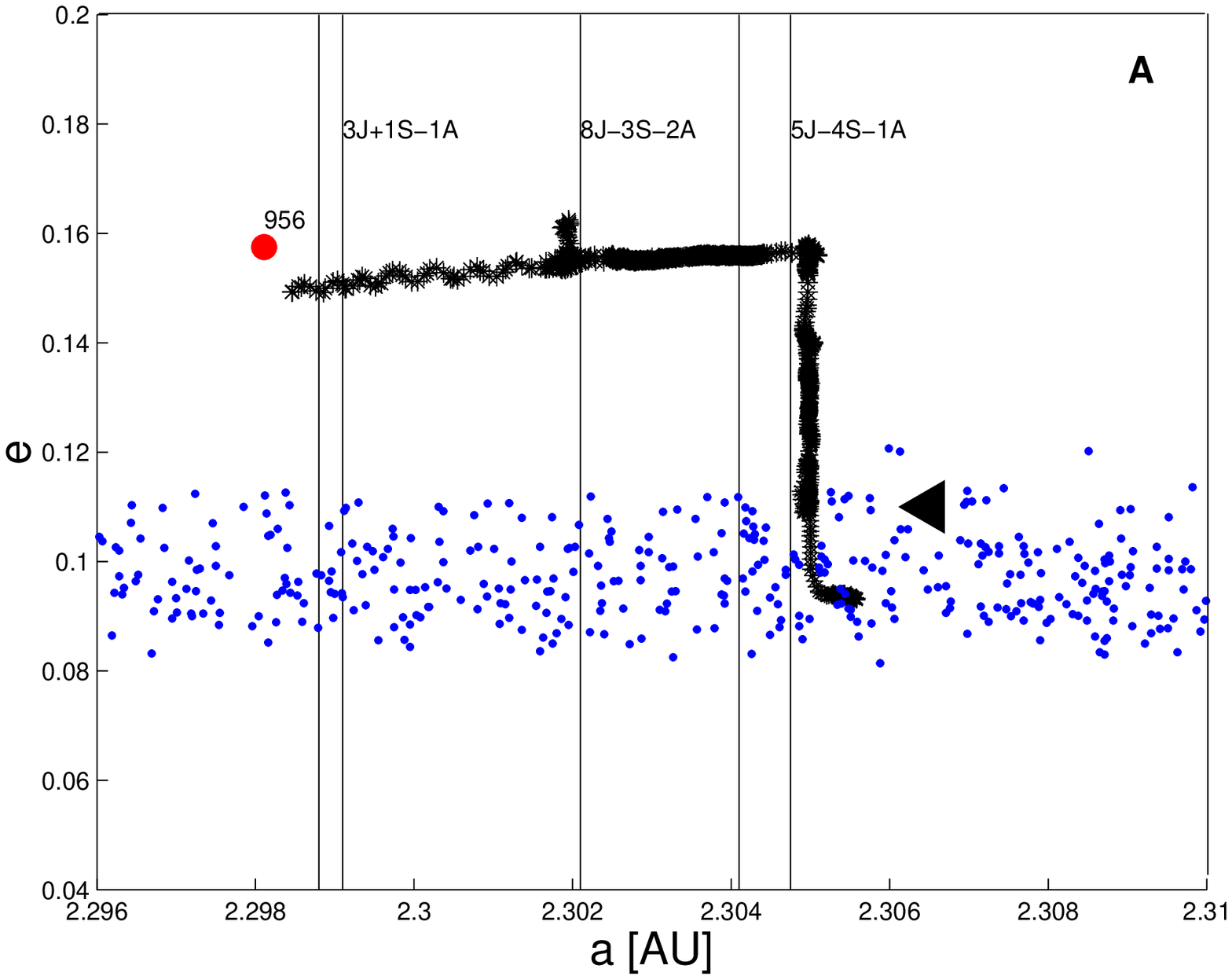}
  \end{minipage}%
  \begin{minipage}[c]{0.5\textwidth}
    \centering \includegraphics[width=2.5in]{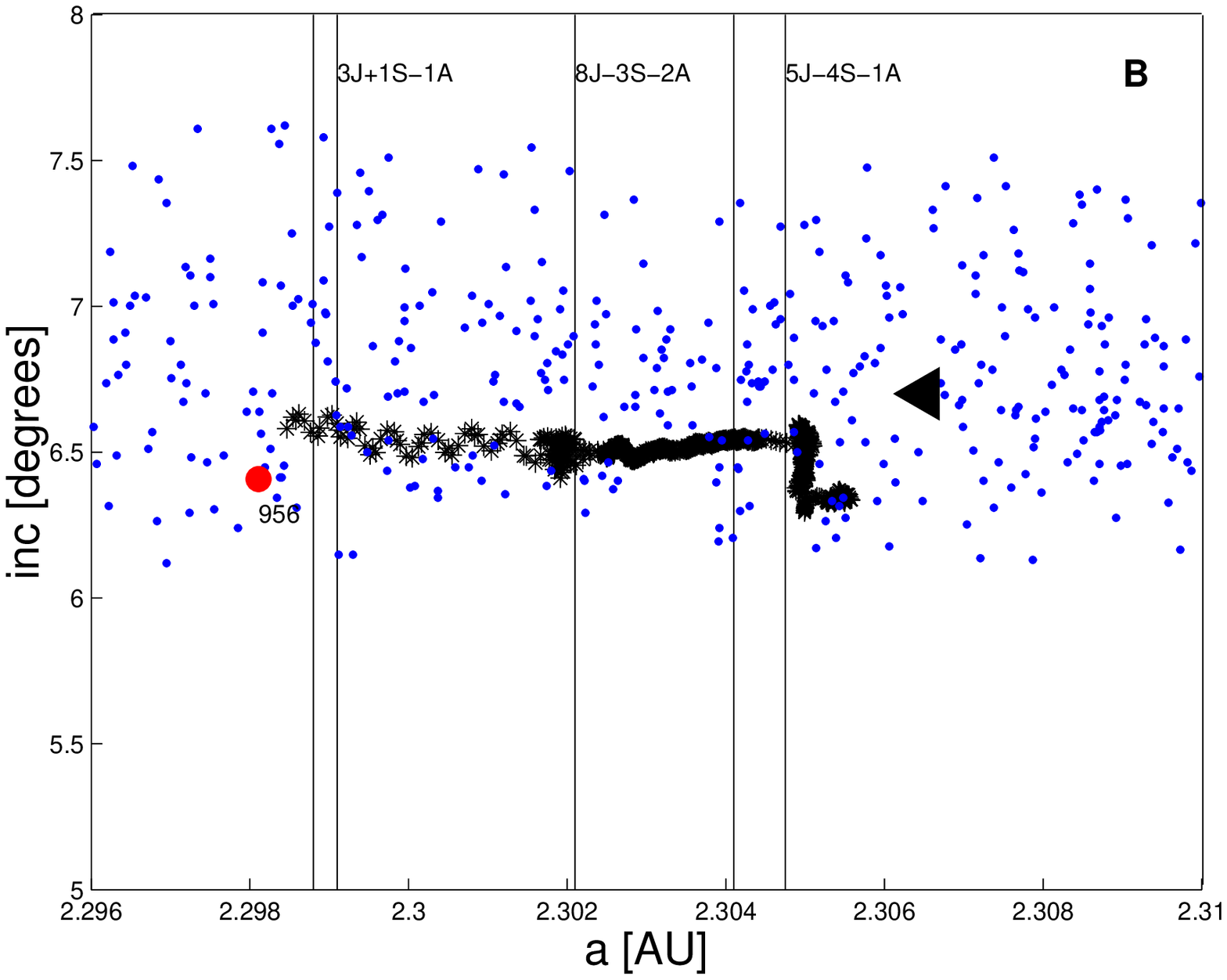}
  \end{minipage}

     \caption{Averaged {\em a-e} (Fig.~\ref{fig: capture-rapture}a), 
and {\em a-i} (Fig.~\ref{fig: capture-rapture}b) evolutions of a 
8000 m radius asteroid near the 5J-4S-1A three-body resonance under 
the effect of the Yarkovsky force.  The black arrows show the directions 
of propagation.  The full dot shows the location (in proper 
element space) of 956 Elisa, while the small dots identify 
Vesta-family members.  Only asteroids with $D$~$>$~8~km can be 
captured in the three-body resonance and remain inside it
for time long enough to 
reach the transition region with the $z_2$ resonance.  Once there, the
particle was captured in the $z_2$ resonance, and drifted until reaching 
values of $a$, $e$, and $i$ comparable to those of 956 Elisa.}
     \label{fig: capture-rapture}
\end{figure}

Fig.~\ref{fig: capture-rapture} shows the {\em a-e} and {\em a-i} 
evolutions of a 8 km clone integrated
with SWIFT-RMVSY (same thermal elements as in Section 3) 
for 1.2 Gyr with initial $e$ equal to 0.09. Of the integrated particles,  
only those with $R$~$>$ 4 km were captured into the 
three-body resonance.  Of the captured particles,
only those with $R$~$=$~$8$~km drifted along the 5J-4S-1 until they 
reached the $z_2$ secular resonance where 
they evolved until reaching the orbital location of 956 Elisa.  These 
results did not significantly depend on the initial values of $e_0$.

This simulation demonstrated that it is possible to be captured 
into the $z_2$ from the 
5J-4S-1 resonance, but that only relatively large bodies 
can stay inside the 5J-4S-1 for times long enough to reach the transition
region with the $z_2$ resonance.  Smaller bodies are either drifting too 
fast to be captured into the 5J-4S-1A resonance or, even if captured, do not 
remain inside this resonance long enough to reach the domain of the $z_2$ 
resonance.  Moreover, if we assume that 956 Elisa is a runaway fragment
of the original Vesta family, this simulation provides an estimate 
on the age of the Vesta family, which should be at least as old as the
time required for Elisa to migrate to its current position (and 
possibly older, since we neglected reorientations, and the evolution
in the three-body and secular resonances was the fastest possible).

To further study how the capture mechanism 
depends on the asteroid size, we integrated 33 particles 
in the semimajor axis range between 2.30360 and 2.30520 AU, with the 
other orbital elements of 956 Elisa (the 5J-4S-1A resonance is limited to the 
interval 2.30410-2.30475 AU).  We made three runs with 
particles of 100 m, 4000 m, 8000 m of radii and with the usual thermal 
parameters.  We want to understand if capture into the $z_2$ depends on the 
asteroid size, and, since we now know that migration in the 5J-4S-1A 
is possible, we started with particles as close as possible to the 
transition region of the 5J-4S-1A with the $z_2$ resonance.

Of the integrated particles inside the 5J-4S-1A resonance, 
58\% of the 100 m, 55\% of the 4000 m, and 
64\% of the 8000 m particles were captured into the $z_2$.  This shows that
capture from the three-body resonance into the $z_2$ is not strongly
dependent on the asteroid size.
The size-dependent selection process, in our opinion, 
operates in the region where the three-body resonances
cross the Vesta dynamical region. In our scenario only relatively 
large asteroids are drifting slow enough via Yarkovsky effect to be captured 
into the 5J-4S-1A resonance, where they evolve until reaching the $z_2$ 
resonance, and are finally captured.  This scenario could explain 
why the three asteroids in the $z_2$ resonance all have estimated diameters
much larger than the average diameter of Vesta family members ($>$10~km, 
as opposed to $\simeq$4~km). 

\subsection{Other nonlinear secular resonances: the ${\gamma}_2$ 
resonance}

In section 3 we showed that asteroids currently inside the $z_2$ resonance 
can interact with other secular resonances when the Yarkovsky effect is 
considered.  For example, a 4000 m clone of 956 Elisa was captured in 
the ${\gamma}_2$ secular resonance.  Here we try to understand if a 
Vesta-family asteroid migrating via the Yarkovsky effect could be 
captured into one of the 
secular resonances and then drift until it meets the $z_2$.
Since our study focuses on the dynamical evolution of 956 Elisa, we 
concentrate on the ${\gamma}_2$ resonance.  Once again, we believe 
that our results may be qualitatively extended to the cases of other 
nonlinear resonances.

We started by simulating fictitious members of the Vesta family 
already inside the ${\gamma}_2$ resonance, and evolving 
under the influence of the 
Yarkovsky effect.  We used 33 clones of 956 Elisa (see 
Fig.~\ref{fig: Elisa-dyn}) with semimajor axis ranging from 2.29360 to 
2.30960 AU, with $e$ equal to 0.19960 and the other elements the 
same as 956 Elisa.  Of these test particles, 10 were in resonance and 10 
more in quasi-resonance.  We made three integrations with particles
with radii of 100 m, 4000 m, and 8000 m, under the influence of the 
Yarkovsky force (we used the usual thermal parameters of V-type asteroids).

Of our simulated particles, the smaller ones with 100 m radius 
left the ${\gamma}_2$ resonance as soon the simulation 
started.  Of the larger particles, only a test particle of 8000 m radius 
stayed inside the ${\gamma}_2$ over the 300 Myr of the integration 
(the integration time, however, was not 
long enough to allow the particle to reach the transition region with the 
$z_2$ resonance).  While it is possible that particles
captured in the ${\gamma}_2$ resonance could migrate until they reach
the $z_2$ resonance, our simulations suggest that this mechanism is 
much less effective than transport via a three-body resonance.

Once test particles reach the transition region, is capture into 
the $z_2$ resonance a size-dependent process? 
To answer this question, we integrated 33 clones of 956 Elisa with semimajor
axes going from 2.29360 to 2.30960, an eccentricity value of 0.12460, and all 
the other elements the same as 956 Elisa, under the effect of the 
Yarkovsky force.  Seven of these clones were initially in the ${\gamma}_2$ 
resonance with the other 8 in semi-resonance.  We made a series 
of three simulations with
particles of 100, 4000, and 8000 m in radius, and the results were always
the same: except for the occasional particle interacting with a three-body 
resonance, all of the simulated asteroids were captured into the $z_2$ 
resonance once they reached the transition region.  Once again, the 
process of capture into the $z_2$ resonance seems not to be a size-dependent
process.

\subsection{The 7J-2A mean-motion resonance}

We saw in section 3 that the clones of the resonant asteroids evolved
inside the $z_2$ until reaching the 7J-2A mean-motion resonance.
After passing through the 7J-2A resonance, none of the integrated particles
remained in the $z_2$ secular resonance.  To understand if it is possible 
to pass from the 7J-2A to the $z_2$, we integrated 28 retrograde 
particles with zero obliquity with SWIFT-RMVSY.  The particles had seven 
values of osculating $e$ (from $e$ = 0.06 to 0.12, with a step of 0.1) and 
an initial {\em a} of 2.26013 AU.  The initial values of the 
angles ($i, \Omega, \omega, M$) were those of 809 Lundia, and we used 
four values of the radii (100 m, 1000m, 4000 m, and 8000 m),  while 
the thermal parameters are 
the same as used in previous simulations.  The particles drifted via 
the Yarkovsky effect to the separatrix of 7J-2A resonance, and then 
continued their evolution.

Of the integrated particles, some of them evolved in the 7J-2A resonance until 
reaching Mars-crossing orbits, others interacted with the 
9J-5S-2A three-body
resonance  but, as expected, none were captured by the $z_2$ resonance.  
While our simulations do not exclude that such a path is possible, it seems
to be quite unprobable.

\section{Discussions}

In this work we studied a possible migration route from the Vesta dynamical
family to the orbits of two V-type asteroids inside the $z_2$
nonlinear secular resonance.  We found that:

\begin{enumerate}
  \item Members of the Vesta dynamical family migrating via the 
Yarkovsky effect are first captured into three-body (5J-4S-1A, 3J+1S-1A, 
etc.), nonlinear secular resonances 
($2(g-g_6) + g_5 - g_4$) and start migrating 
towards orbits of higher eccentricities.  Only asteroids of $D$~$>$~8~km 
can be captured in one of these resonances and this could explain 
the fact that the three V-type asteroids in the $z_2$ have estimated diameters 
larger than 10 km, which is above the average size of Vesta family members.

   \item Once they reach the zone where those resonances cross
the $z_2$ secular resonance, members of the Vesta family can be 
captured and start their evolution towards larger or smaller 
$a$ inside the resonance.  The process of capture into the $z_2$ resonance 
is essentially size-independent.

   \item Since the typical times required for the clones of 
Elisa to migrate from the Vesta family to the current location of 956 
Elisa is of at least 1.2 Gyr, if we assume that 
956 Elisa was formed in the same cratering event 
that created the Vesta family, our simulations 
set a lower limit on the age of the family.

   \item Asteroids generally stay inside the $z_2$ secular 
resonance until they reach relatively strong mean motion resonances like 
the 7:2 (or other weaker three- or four- body mean motion resonances).
The residence time inside the $z_2$ is of the order of 1 Gyr.

   \item We believe other V-type asteroids could have followed the same path 
and currently be in the $z_2$ secular resonance.  We identify 
several possible V-type asteroids inside this resonance.

\end{enumerate}

While in this paper we suggested a possible scheme for the origin of two
V-type asteroids outside the Vesta dynamical family, currently 20 other 
similar bodies are known.  The migration of these bodies from the Vesta 
family may have involved evolution in other nonlinear secular resonances, 
or different mechanisms of dynamical migration, like 
gravitational scattering by massive asteroids (Carruba {\em et al.} 2003).  
We believe this is an interesting problem, that  
promises to offer still several challenges to dynamicists.

\begin{figure}
  \centering
   \includegraphics[width=0.75\textwidth]{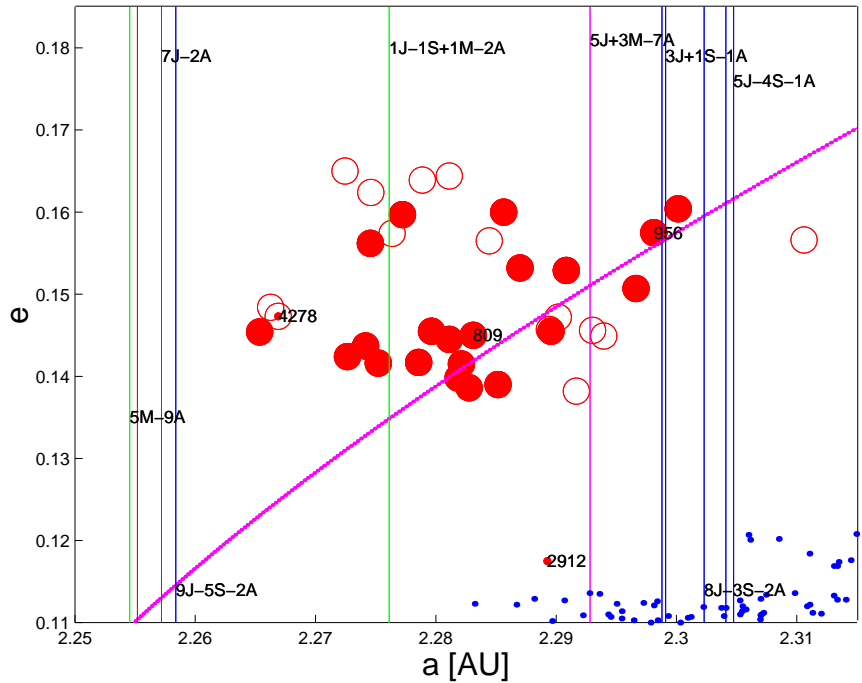}
     \caption{Neighbors of 809 Lundia and 956 Elisa 
obtained with  hierarchical clustering method for an 
absolute magnitude of 14, and a cutoff 
velocity of 130 m/s (Zappal\`{a} {\em et al.} 1990). The asteroids 
currently in the $z_2$ resonance are indicated by full dots, while the 
non-resonant ones are shown by empty dots.  The diagonal
line displays the location of the $z_2$ secular resonance, computed for the 
inclination of 4 Vesta, in osculating $a-e$ space.  Small dots show the 
orbits of family members.  Note how 
4278 Harvey could have been found by using this method.  }
     \label{fig: res-neigh}
\end{figure}

On the other hand, now that we know it is possible to 
migrate from the Vesta family to 
the $z_2$ secular resonance we may ask ourselves if other asteroids may have 
followed the same route.  Are there any other V-type asteroids
near 809 Lundia and 956 Elisa?
To answer this question we searched for 
minor bodies close in proper element space to 
the three resonant asteroids.  We only considered objects of absolute 
magnitude up to 14.5 (the current maximum magnitude of V-type asteroids 
outside the Vesta family in the inner belt) and with mutual distances in 
proper element space of less than 130 m/s.  Table~\ref{table: V-ast-sg4}
and Fig.~\ref{fig: res-neigh} show the asteroids we found.  With the 
exception of 4278 Harvey (and, of course, the two resonant asteroids),
that has a V-type spectrum, no spectral classification is currently 
available for the other bodies.

We integrated these objects for 6 Myr and verified that most of them 
are currently inside the $z_2$ resonance.
In our opinion, some of these asteroids
could have migrated from the Vesta 
family in a way similar to that of 809 Lundia and 956 Elisa.   
We encourage observers to obtain a classification for these
bodies in order to determine if any of them 
belong to the V-type class.

\begin{acknowledgements}

We are grateful to Dr. Robert Jedicke for the 
careful revision of our paper and for helpful comments. Part of 
this work was supported by FAPESP (grant 03/07462-8) and CNPq.  

\end{acknowledgements}

\newpage

\begin{table}
\begin{center}
\caption{Secular resonances involving the perihelion, the 
node, and combinations of perihelion and node of the asteroid.  
The first, third, and fifth columns report the resonant argument 
in terms of the planetary frequencies, 
and the second, fourth, and sixth we give the names we use in this article.}
\label{table: sec-per-nod}
\vspace{0.5cm}
\begin{tabular}{|c|c|c|c|c|c|}
\hline
                             &         &            &       &                         &  \\
Perihelion res.              &  Names  &  Node res. & Names & Perihelion and node res.& Names\\
                             &         &            &       &                         &  \\
\hline
                             &     &                        &     &                            &      \\
$g - 2g_5 - g_6 + 2g_7$      & ${\gamma}_1$     &  $s -s_6 + g_6 - g_7$  & ${\sigma}_1$    & $2g - g_6 - g_4 + s - s_4$ & ${\psi}_1$  \\
$2g - 2g_6 +g_5 - g_4$       & ${\gamma}_2$     &  $s -s_6 - g_5 + g_6$  & ${\sigma}_2$    & $2g - g_6 - g_7 + s - s_7$ & ${\psi}_2$  \\
$2g - 2g_6 + g_7 - g_4$      & ${\gamma}_{3a}$  &  $s - 2s_6 + s_7$      & ${\sigma}_3$    & $2g - g_5 - g_6 + s - s_7$ & ${\psi}_3$  \\
$2g - 3g_6 + g_4$            & ${\gamma}_{3b}$  &  $s - 2s_6 + s_4$      & ${\sigma}_4$    & $2(g-g_6) + s - s_6$       & ${\psi}_4$,$z_2$\\
$g + 2g_5 - 3g_4$            & ${\gamma}_4$     &  $s - s_6 - g_7 + g_4$ & ${\sigma}_{5a}$ & $g - g_4 + 2s - 2s_6$      & ${\psi}_5$  \\
$g + g_5 + g_7 - 3g_4$       & ${\gamma}_{5a}$  &  $s - s_6 + g_6 - g_4$ & ${\sigma}_{5b}$ & $g - g_5 + 2s- s_6 - s_4$  & ${\psi}_6$  \\
$g + g_5 - g_6 - g_4$        & ${\gamma}_{5b}$  &  $s - s_4 - g_5 + g_6$ & ${\sigma}_{6}$  & $g - g_7 + 2s - s_6 - s_4$ & ${\psi}_7$  \\
$g + g_5 - 2g_6 - g_7 + g_4$ & ${\gamma}_{5c}$  &  $s -s_4 + g_6 - g_7$  & ${\sigma}_7$    & $g - g_5 + 2s- 2s_6$       & ${\psi}_8$  \\
$g + 2g_7 - 3g_4$            & ${\gamma}_{6a}$ &                        &                 & $g - g_7 +2s - 2s_6$       & ${\psi}_9$  \\
$g - g_6 + g_7 - g_4$        & ${\gamma}_{6b}$  &                        &                 &                            &      \\
$g - 2g_6 + g_4$             & ${\gamma}_{6c}$  &                        &                 &                            &      \\
                             &     &                        &     &                            &      \\
\hline
\end{tabular}
\end{center}
\end{table}

\newpage

\begin{table}
\begin{small}
\begin{center}
\caption{List of the 22 V-type asteroids currently known outside 
the Vesta dynamical family.  We report number, name, proper 
{\em a,e, i}, absolute magnitude H, if they are in (R, which stands 
for resonant) or close (C, for Circulating) to a nonlinear 
secular resonance (see codes in table~\ref{table: sec-per-nod}), and, 
if they are close to one of the resonances, what is 
the period of circulation of the resonant
argument.}
\label{table: V-ast}
\vspace{0.02cm}
\begin{tabular}{|c|c|c|c|c|c|c|c|}
\hline
        &           &         &        &        &       &          &                    \\
 Ast. \# & Ast. Name & a       &  e     & i      &  H    & Sec. Res.& Circulation Period \\
        &           &         &        &        &       &          &                    \\
\hline
        &           &         &        &        &       &          &                    \\
809  & Lundia       & 2.28311 & 0.1450 & 6.7363 & 12.10 & $z_2$    & R             \\
956  & Elisa        & 2.29811 & 0.1575 & 6.4076 & 12.03 & $z_2$,${\gamma}_2$ & R, C $>$ 4 Myr\\

2113 & Ehrdni       & 2.47377 & 0.0726 & 6.0790 & 12.66 &        &                       \\
2442 & Corbett      & 2.38765 & 0.0972 & 5.4225 & 12.54 & ${\sigma}_7$     & C $>$ 1 Myr        \\
2566 & Kirghizia    & 2.45003 & 0.1044 & 4.4219 & 12.20 & ${\sigma}_7$, 9J-7S-2A & C $>$ 4 Myr, C $>$ 0.05 Myr \\
2579 & Spartacus    & 2.21037 & 0.0807 & 5.8947 & 13.03 &        & \\
2640 & Hallstrom    & 2.39777 & 0.1253 & 6.5286 & 12.85 & 4J-2S-1A, ${\sigma}_6$, ${\sigma}_7$ & R, C $>$ 4 Myr,C $>$ 4 Myr,\\
2653 & Principia    & 2.44350 & 0.1145 & 5.0773 & 12.24 &        & \\     
2704 & Julian Loewe & 2.38485 & 0.1173 & 5.0888 & 12.75 & ${\sigma}_7$     & C $>$ 1 Myr        \\
2763 & Jeans        & 2.40355 & 0.1790 & 4.3185 & 11.99 &        & \\
2795 & Lepage       & 2.29574 & 0.0747 & 6.5979 & 12.75 & ${\gamma}_2$     & C $>$ 4 Myr        \\
2851 & Harbin       & 2.47819 & 0.1204 & 7.7933 & 11.98 & ${\gamma}_{5a}$    & C $>$ 3 Myr        \\
2912 & Lapalma      & 2.28927 & 0.1175 & 6.7421 & 12.45 & ${\gamma}_7$     & C $>$ 1 Myr        \\
3849 & Incidentia   & 2.47422 & 0.0653 & 5.3938 & 12.63 & ${\gamma}_{5b}$, ${\gamma}_{6b}$ & C $>$ 5 Myr,C $>$ 5 Myr \\
3850 & Peltier      & 2.23457 & 0.1061 & 4.7323 & 13.46 & ${\sigma}_4$     &  C $>$ 4 Myr \\
3869 & Norton       & 2.45241 & 0.1009 & 5.2787 & 12.54 &        &                 \\
4188 & Kitezh       & 2.33538 & 0.1118 & 5.6413 & 12.67 & ${\sigma}_{5a}$    &  C $>$ 3 Myr \\
4278 & Harvey       & 2.26689 & 0.1473 & 5.2499 & 13.62 & ${\psi}_2$, ${\sigma}_{5b}$ & C $>$ 2 Myr, C $>$ 6Myr         \\ 
4434 & Nikulin      & 2.44110 & 0.1033 & 5.5261 & 12.88 & ${\gamma}_{6c}$    &  C $>$ 2 Myr \\
4796 & Lewis        & 2.35523 & 0.1413 & 3.0955 & 13.22 & ${\sigma}_6$, ${\sigma}_7$  &  C $>$ 4 Myr,C $>$ 4 Myr \\
4977 & Rauthgundis  & 2.29229 & 0.0928 & 5.9408 & 13.99 & ${\gamma}_2$     & C $>$ 2 Myr        \\
5379 & Abehiroshi   & 2.39599 & 0.0728 & 3.3537 & 12.55 & ${\gamma}_{5c}$, ${\gamma}_{6c}$ & C $>$ 4 Myr,C $>$ 4 Myr \\
        &           &         &        &        &       &          &                    \\
\hline
\end{tabular}
\end{center}
\end{small}
\end{table}

\begin{table}
\begin{small}
\begin{center}
\caption{The 33 asteroids close in proper element space to 809 Lundia,
and 956 Elisa, identified with the hierarchical clustering method 
described in Section 4. We report the 
asteroid identification, the proper $a,e,sin{(i)}$, if the asteroid is 
inside the $z_2$ secular resonance, and the spectral type (if known).}
\label{table: V-ast-sg4}
\vspace{0.5cm}
\begin{tabular}{|c|c|c|c|c|c|}
\hline
        &       &          &           &            &             \\
Ast. \# & $a$   & $e$      & $\sin(i)$ & $z_2$ res. & Spect. type \\
        &       &          &           &            &             \\
\hline
        &       &          &           &            &             \\
956   & 2.29812 & 0.1575 & 0.1116 & yes & V \\
11918 & 2.3106  & 0.1566 & 0.114  & no  &   \\
24840 & 2.30014 & 0.1604 & 0.1131 & yes &   \\
7589  & 2.28564 & 0.16   & 0.1135 & yes &   \\
3083  & 2.28444 & 0.1565 & 0.1168 & no  &   \\
14948 & 2.27722 & 0.1597 & 0.1107 & yes &   \\
16556 & 2.28112 & 0.1644 & 0.1129 & no  &   \\
17251 & 2.287   & 0.1532 & 0.1132 & yes &   \\
5238  & 2.27456 & 0.1562 & 0.1105 & yes &   \\
22351 & 2.27637 & 0.1574 & 0.1095 & no  &   \\
6481  & 2.27887 & 0.1639 & 0.1161 & no  &   \\
20070 & 2.27245 & 0.165  & 0.1099 & no  &   \\
25343 & 2.2746  & 0.1624 & 0.1162 & no  &   \\
16377 & 2.29084 & 0.1529 & 0.1136 & yes &   \\
24755 & 2.29664 & 0.1507 & 0.1156 & yes &   \\
        &       &          &           &            &             \\
\hline 
        &       &          &           &            &             \\
809   & 2.28311 & 0.145  & 0.1173 & yes & V \\
4278  & 2.26689 & 0.1473 & 0.0915 & no  & V \\
21829 & 2.27416 & 0.1437 & 0.1171 & yes &   \\
12227 & 2.27266 & 0.1424 & 0.114  & yes &   \\
18082 & 2.26536 & 0.1454 & 0.1163 & yes &   \\
21935 & 2.27521 & 0.1416 & 0.1179 & yes &   \\
22598 & 2.27857 & 0.1417 & 0.1128 & yes &   \\
19373 & 2.26626 & 0.1484 & 0.1146 & no  &   \\
19738 & 2.28213 & 0.1415 & 0.121  & yes &   \\ 
1476  & 2.28112 & 0.1445 & 0.1124 & yes &   \\
24918 & 2.28185 & 0.1398 & 0.1151 & yes &   \\
20181 & 2.28277 & 0.1386 & 0.1222 & yes?&   \\
17453 & 2.28957 & 0.1455 & 0.1118 & yes &   \\
19672 & 2.28517 & 0.139  & 0.1168 & yes?&   \\  
12671 & 2.29168 & 0.1382 & 0.1198 & no  &   \\
14191 & 2.29302 & 0.1456 & 0.1089 & no  &   \\
7550  & 2.29017 & 0.1472 & 0.1079 & no  &   \\
16740 & 2.28944 & 0.1457 & 0.1066 & no  &   \\
24303 & 2.29396 & 0.1449 & 0.1062 & no  &   \\
14444 & 2.27964 & 0.1455 & 0.1077 & yes &   \\
        &       &          &           &            &             \\
\hline
\end{tabular}
\end{center}
\end{small}
\end{table}

\end{document}